\documentclass[conference]{IEEEtran}
\IEEEoverridecommandlockouts
\usepackage{cite}
\usepackage{amsmath,amssymb,amsfonts}
\usepackage{algorithmic}
\usepackage{graphicx}
\usepackage{textcomp}
\usepackage{xcolor}
\def\BibTeX{{\rm B\kern-.05em{\sc i\kern-.025em b}\kern-.08em
    T\kern-.1667em\lower.7ex\hbox{E}\kern-.125emX}}
    
\usepackage{subcaption}
\usepackage{amsmath,graphicx}
\usepackage{multirow}
\usepackage{amssymb}
\usepackage{mathtools,amsthm}
\usepackage{algorithm}
\usepackage{hyperref}
\usepackage{epstopdf}
\usepackage{amsthm}
\usepackage{comment}
\usepackage{enumitem}
\usepackage{cite}
\usepackage{lipsum}
\usepackage{balance}
\usepackage[algo2e]{algorithm2e}
\usepackage[numbers]{natbib}
\usepackage{times}
\usepackage{makecell}
 \usepackage{soul}
\soulregister\cite7
\soulregister\ref7
\soulregister\pageref7
\usepackage{diagbox}
\usepackage[export]{adjustbox}
\renewcommand{\footnotesize}{\scriptsize}
\usepackage{bm}

\newtheorem{mydef}{Definition}
\newtheorem{lemma}{Lemma}

\newtheorem{theorem}{Theorem}[section]

\usepackage{soul}
\usepackage{comment}
\usepackage{caption}
\usepackage[font=footnotesize]{subcaption}

\usepackage{cleveref}

\newlength\myindent
\usepackage{multirow,tabularx}

\begin{document}

\title{User-Entity Differential Privacy \\ in Learning Natural Language Models
\thanks{$^*$Corresponding author}
}

\author{\IEEEauthorblockN{Phung Lai, NhatHai Phan$^*$}
\IEEEauthorblockA{
\textit{New Jersey Institute of Technology, USA}\\
\{tl353, phan\}@njit.edu}
\and
\IEEEauthorblockN{Tong Sun, Rajiv Jain, Franck Dernoncourt, \\Jiuxiang Gu, Nikolaos Barmpalios}
\IEEEauthorblockA{
\textit{Adobe Systems Inc., USA} \\
\{tsun, rajijain, dernonco, jigu, barmpali\}@adobe.com}
}

\IEEEoverridecommandlockouts
\IEEEpubid{\makebox[\columnwidth]{978-1-6654-8045-1/22/\$31.00~\copyright2022 IEEE\hfill}
\hspace{\columnsep}\makebox[\columnwidth]{ }}

\maketitle
\IEEEpubidadjcol

\begin{abstract}
In this paper, we introduce a novel concept of user-entity differential privacy (UeDP) to provide formal privacy protection simultaneously to both sensitive entities in textual data and data owners in learning natural language models (NLMs). To preserve UeDP, we developed a novel algorithm, called UeDP-Alg, optimizing the trade-off between privacy loss and model utility with a tight sensitivity bound derived from seamlessly combining user and sensitive entity sampling processes. An extensive theoretical analysis and evaluation show that our UeDP-Alg outperforms baseline approaches in model utility under the same privacy budget consumption on several NLM tasks, using benchmark datasets.
\end{abstract}


\section{Introduction} 
\label{intro}

Despite remarkable performance in many applications, natural language models (NLMs), such as GPT models \citep{radford2018improving,radford2019language,brown2020language}, are vulnerable to privacy attacks because of such attacks’ capacity to memorize unique patterns in training data \citep{carlini2019secret}. Recent data training extraction attacks \citep{carlini2021extracting} illustrate that sensitive entities, such as a person’s name, email address, phone number, physical address, etc., can be accurately extracted from NLM parameters. These sensitive entities and the language data memorized in NLMs may identify a data owner - explicitly by name or implicitly, e.g., via a rare or unique phrase - and link that data owner to extracted sensitive entities.

Our main goal is to provide a rigorous guarantee that a trained NLM protects the privacy of sensitive entities in the training data and the participation information (membership) of the data owners in learning the model while maintaining high model utility. The simple solution of anonymizing (including removing/de-identifying) sensitive entities is insufficient; since the anonymized 
entities can be matched with non-anonymized data records in another dataset \citep{dwork2014algorithmic}. Also, the model utility can be notably affected, as shown in our experimental study. While cryptographic approaches can be applied to protect privacy, they introduce computation and resource overhead \citep{al2020privft}. Therefore, we proposed to use differential privacy \citep{dwork2006calibrating}, one of the adequate solutions, given its formal  protection without undue sacrifice in computation efficiency and model utility.

Differential privacy (DP) provides rigorous privacy protection as a probabilistic term, limiting the knowledge about a data record an ML model can leak while learning features of the whole training set. DP-preserving mechanisms have been investigated and applied in real-world \citep{abadi2016deep,phan2016differential,shokri2015privacy}, including image processing  \citep{phan2020scalable}, healthcare data \citep{zia2020application}, financial records \citep{wu2019value}, social media \citep{li2012sampling}, and NLMs \citep{mcmahan2017learning, lyu2020towards,lyu2020differentially,bagdasaryan2019differential}.

However, existing DP protection levels, including sample-level DP \citep{abadi2016deep, roth2012buying, dwork2014algorithmic,bassily2014private}, user-level DP \citep{mcmahan2017learning, ramaswamy2020training}, element-level DP \citep{asi2019element}, and local (feature-level) DP \citep{lyu2020differentially, lyu2020towards, erlingsson2014rappor,duchi2018minimax}, do not provide the privacy protection level demanded to solve our problem. Given training data: 1) Sample-level DP protects the privacy of a single sample; 2) User-level DP protects privacy of a single data owner, also called a single user, who may contribute one or more data samples; 3) Element-level DP partitions data owners' contribution to the training data into sensitive elements, e.g., a curse word, which will be protected. Element-level DP does not provide privacy protection to data owners; and 4) Local (feature-level) DP protects true values of a data sample from being inferred. 
Recently, \citep{lyu2020differentially} proposed local DP-preserving approaches for text embedding extraction under (word-level) local  DP (\textbf{Eq. \ref{word-level LDP}}). 
However, the privacy budget in \citep{lyu2020differentially} is accumulated over the dimensions of embedding, resulting in an impractical (loose) privacy guarantee (Appendix \ref{lyu2020differentially} in our supplemental document\footnote{\url{https://www.dropbox.com/s/3ch7m2yskshkkc5/UeDP_Supplementary.pdf?dl=0}}).


Therefore, there is a demand for a new level of DP to protect privacy simultaneously for both sensitive entities in the training data and the participation information of data owners in learning NLMs.
Motivated by this, we structure our paper around the following significant contributions. 

$\bullet$ We propose a novel notion of user-entity adjacent databases (\textbf{Definition \ref{UeAB}}), leading to formal guarantees of user-entity privacy rather than privacy  for a single user or a single sensitive entity.

$\bullet$ To preserve UeDP, we introduce a novel algorithm, called \textbf{UeDP-Alg}, which leverages the recipe of \textsc{DP-FedAvg} \citep{mcmahan2017learning} to protect both sensitive entities and user membership under DP via the moments accountant \citep{abadi2016deep}. Moments accountant was first developed to preserve DP in stochastic gradient descent (SGD) for sample-level privacy. Our federated averaging approach groups multiple SGD updates computed from a two-level random sampling process, including a random sample of users and a random sample of sensitive entities. That enables large-step model updates and optimizes the trade-off between privacy loss and the model utility through a tight noise scale bound (\textbf{Lemma \ref{lemma2}} and \textbf{Theorem \ref{theorem1}}).

$\bullet$ Through theoretical analysis and rigorous experiments conducted on benchmark   datasets, we show that our UeDP-Alg outperforms baseline approaches in terms of model utility on fundamental tasks, i.e., next word prediction and text classification, under the same privacy budget consumption. Our code is available\footnote{\url{https://github.com/PhungLai728/UeDP}}.


\section{Background} 
\label{background}


In this section, we revisit NLM tasks, privacy risk, and DP. For the sake of clarity, let us focus on the next word prediction, and we will extend it to text classification in Section \ref{exp}.   A list of sensitive entity categories is summarized in Table~\ref{tb3}.

\paragraph{Next Word Prediction} Let $D$ be a private training data containing $U$ users (data owners) and a set of sensitive entities $E$. Each user $u\in U$ consists of $n_u$ sentences. 
Given a vocabulary $\mathcal{V}$, each sentence is a sequence of words, presented as $x = x_1 x_2 \ldots x_{m_u}$, where $x_i \in \mathcal{V}, (i \in [1, m_u]) $ is a word in $x$ and $m_u$ is the length of $x$.
In next word prediction, the first $j$ words in $x$, i.e., $x_1, x_2, \ldots, x_j$ ($\forall j <m_u$), are used to predict the next word $x_{j+1}$. Here, $x_{j+1}$ can be considered as a label in the next word prediction task. Perplexity $PP = 2^{-\sum_{x \in D} p(x) \log_2 p(x)}$ is a measurement of how well a model predicts a sentence and is often used to evaluate language models, where $p(x)$ is a probability to predict the next word $x_{j+1}$ in $x$ \citep{mikolov2011empirical}. 
A lower perplexity indicates a better model. 

\paragraph{Sensitive Entities and Sentences}  
Each sensitive entity $e \in E$ consists of a word or consecutive words that must be protected. 
For instance, personal identifiable information (PII) related to an identifiable person, such as person names, locations, and phone numbers, can be considered sensitive entities. If a sentence $x$ consists of a sensitive entity $e$, $x$ is considered as a sensitive sentence; otherwise, $x$ is a non-sensitive sentence.  

\begin{figure}[t] 
      \centering
      \includegraphics[scale=0.41]{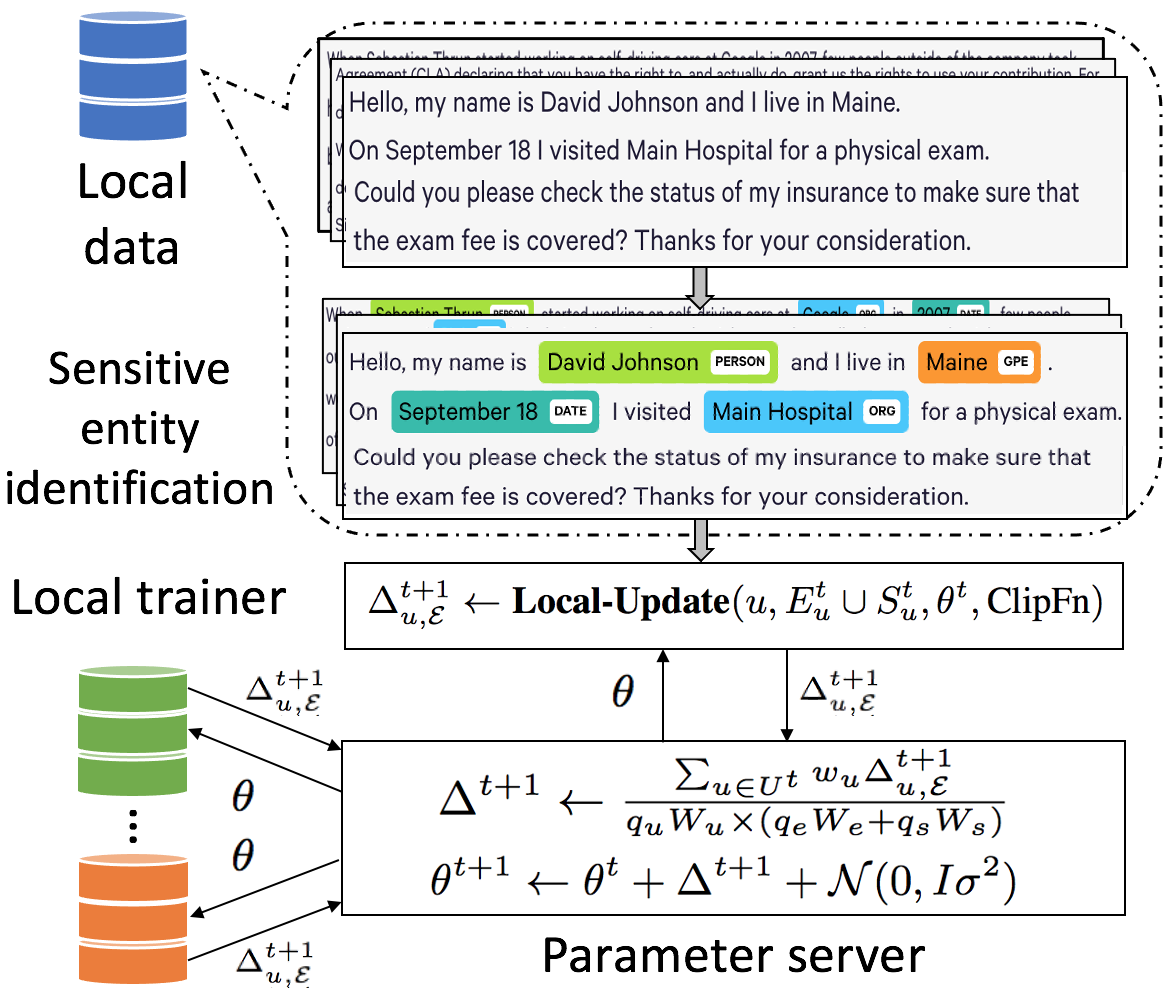} 
      \caption{User-Entity DP. Data from users is processed to identify sensitive entities, before being trained by local trainers. Bounded gradients from local trainers are aggregated at a server with additive noise. Updated model is sent back to local trainers for next rounds.  
      } 
      \label{fig10}
 \end{figure}

For instance, in Fig.~\ref{fig10}, 
``David Johnson,'' ``Maine,'' ``September 18,'' and ``Main Hospital'' are considered sensitive entities, correspondingly categorized into PII, geopolitical entities (GPE) (i.e., countries, cities, and states), time, and organization names. The first and second sentences consisting of the sensitive entities are considered sensitive sentences.
Meanwhile, the third and fourth sentences are non-sensitive since they do not contain any sensitive entities.

\paragraph{Privacy Threat Models}
It is well-known that trained ML model parameters can disclose information about training data 
\citep{carlini2021extracting,dwork2008differential}, 
especially in NLMs \citep{mcmahan2017learning,carlini2021extracting}. Given a data sample and model parameters, by using a membership inference attack \citep{shokri2017membership, salem2018ml,yeom2018privacy}, adversaries can infer whether the training used the sample or not. 
In NLMs, adversaries can accurately recover individual training examples, such as full names, email addresses, and phone numbers of individuals, using training data extracting attacks \citep{carlini2021extracting}. Accessing to these can lead to severe privacy breaches.

\begin{table}[t]
\footnotesize
\caption{Description of sensitive entity categories.}
\label{tb3}
\begin{center}
\begin{tabular}{| l |  l|}
\hline
Type & Description \\
\hline \hline
Person & Person, i.e.,  people, including fictional\\
Loc & Location, i.e., non-GPE locations, mountain ranges, bodies of water \\
Org & Organization, i.e., companies, agencies, institutions, etc. \\
Misc & Miscellaneous, i.e., entities that do not belong to the person, \\ 
& location,  and organization in CONLL-2003  \\
GPE & Geopolitical entity, i.e., countries, cities, states \\
PII & Personal identification information, i.e., name, location, phone, etc.   \\
Date & Absolute or relative dates or periods \\
NoRP & Nationalities or religious or political groups  \\
Fac & Buildings, airports, highways, bridges, etc. \\
Product &  Objects, vehicles, foods, etc. (Not services.) \\
Event & Named hurricanes, battles, wars, sports events, etc.\\
Law & Named documents made into laws \\
Language & Any named language \\
Work of art  & Titles of books, songs, etc. \\
Time &  Times smaller than a day \\
Percent & Percentage, including ``$\%$''  \\
Money & Monetary values, including unit \\
Quantity & Measurements, as of weight or distance \\
Ordinal &  ``First'', ``second'', etc.  \\
Cardinal & Numerals that do not fall under another type  \\
 \hline
\end{tabular} 
\end{center}
\end{table}

\section{Different Levels of DP}

To avoid these privacy risks, DP guarantees restriction on the adversaries in what they can learn from the training data given the model parameters by ensuring similar model outcomes with and without any single training sample.
Let us revisit the definition of DP, as follows:

\begin{mydef}{$(\epsilon, \delta)$-DP \citep{dwork2006calibrating}.} A randomized algorithm $\mathcal{A}$ fulfills $(\epsilon, \delta)$-DP, if for any two adjacent datasets $D$ and $D'$ differing by at most one sample, and for all outcomes $\mathcal{O} \subseteq Range(\mathcal{A})$:
\begin{equation}
Pr[\mathcal{A}(D) = \mathcal{O}] \leq e^{\epsilon} Pr[\mathcal{A}(D') = \mathcal{O}] + \delta 
\end{equation}
with a privacy budget $\epsilon$ and a broken probability $\delta$. 
\label{Different Privacy} 
\end{mydef}

The privacy budget $\epsilon$ controls the amount by which the distributions induced by $D$ and $D'$ may differ. A smaller $\epsilon$ enforces a stronger privacy guarantee.
The broken probability $\delta$ means the highly unlikely ``bad'' events, in which an adversary can infer whether a particular data sample belongs to the training data, happen with the probability $\le \delta$.

There are different levels of DP protection in literature categorized into four research lines, including sample-level DP, user-level DP, element-level DP, and local (feature-level) DP. They are different from our goal since we focus on providing simultaneous protections to data owners and sensitive entities in textual data.  
Let us revisit these DP levels and distinguish them with our goal.

\paragraph{Sample-level DP}
Traditional DP mechanisms  \citep{roth2012buying, dwork2014algorithmic, pan2020privacy} ensure DP at the sample-level, in which adjacent datasets $D$ and $D'$ are different from at most a single training sample. 
Sample-level DP does not protect privacy for users. That is different from our goal. We aim at protecting privacy for users and sensitive entities, which are different from data samples. 




\paragraph{User-level DP} To protect privacy for users, who may contribute more than one training sample, rather than a single sample, 
\citep{mcmahan2017learning} proposed a user-level DP, in which neighboring databases $D$ and $D'$ are defined to be different from all of the samples associated with an arbitrary user in the training set. 
Several works follow this direction \citep{kairouz2019advances,ramaswamy2020training}. 
User-level DP differs from our goal, since it does not provide privacy protection for sensitive entities in the training set. 





\paragraph{Element-level DP} \citep{asi2019element} introduce element-level DP, in which users are partitioned based on sensitive elements, which are protected in a way that an adversary cannot infer whether a user has a sensitive element in her/his data, e.g., if a user has ever sent a curse word in his/her messages or not. 
Similar to sample-level DP, element-level DP is different from our goal, since it does not provide DP protection for users. 


\paragraph{Local (feature-level) DP} \citep{lyu2020differentially} proposed a notion of word-level local DP for a sentence's embedding features, in which two adjacent sentences $x$ and $x'$ are different at most one word: 
\begin{equation}
Pr[\mathcal{A}(f(x)) = \mathcal{O}] \leq e^\epsilon Pr[\mathcal{A}(f(x')) = \mathcal{O}]
\label{word-level LDP}
\end{equation}
where $f(x)$ extracts embedded features of $x$ and $\mathcal{A}$ is a randomized algorithm, such as a Laplace mechanism \citep{dwork2014algorithmic}.
In a similar effort, \citep{lyu2020towards} applied a randomized response mechanism \citep{erlingsson2014rappor,wang2017locally,arachchige2019local} on top of binary encoding of embedded features' real values to achieve local DP feature embedding.
The approaches proposed in \citep{lyu2020differentially,lyu2020towards} are different from our goal, since they do not offer either user-level DP or word-level $\text{DP}$.

\section{User-Entity Differential Privacy}
\label{uedp}


In this section, we focus on answering the question: \textit{``Could we protect sensitive entities and user membership simultaneously by leveraging existing levels of DP and how?''} Based upon that, we propose our user-entity DP notion.

\subsection{Sensitive Entities and User Membership} 

To protect sensitive entities and user membership, a potential approach is to decouple them into separated protection levels offering by existing DP notions. However, this approach has limitations as discussed next.


Let us consider a sentence consisting of one or more than one sensitive entities. We can leverage sample-level DP to protect the sentence, i.e., each sentence could be a sample, covering all the sensitive entities under DP. If each user has only one sentence, then this approach can also protect the user membership. In practice, one user may contribute many sentences to the training data. To address this issue, we can utilize group privacy \citep{dwork2014algorithmic} resulting in an amplification of the privacy budget proportional to the number of sentences a user may have in the training data.  

Instead of group privacy, another potential solution is applying user-level DP on top of the sample-level DP to protect both sentence and user membership. In the sample-level DP, we can clip and inject Gaussian noise into the gradient derived from each sentence \citep{abadi2016deep}. Meanwhile, in the user-level DP, an additional Gaussian noise is injected into the aggregation of gradients, each of which derived from a single user \citep{mcmahan2017learning}. Although this combination of sample - user levels can cover both sensitive entities and user membership under DP protection, it has disadvantages. 
First, some sentences are sensitive and other sentences are not. Protecting all (sensitive and non-sensitive) sentences or removing all the sensitive sentences from the training data may cause significant model utility degradation. Second, different sentences may consist of different types and numbers of sensitive entities. Under the same sampling probability for training as in \citep{abadi2016deep} for sample-level DP, these sentences expose different privacy risks to user identity and sensitive entities. 

To address these issues, instead of the sentence level, one can work at the word level by extracting embedded features for every words in the training data. Embedded features of sensitive entities are randomized by local DP-preserving mechanisms \citep{arachchige2019local}. The randomized embedded features are aggregated with embedded features of non-sensitive words to train NLMs. Then, user-level DP can be applied to clip gradients derived from each user's data with adding Gaussian noise into the aggregation of these gradients. However, this approach suffers from a remarkable model utility degradation. Local DP provides rigorous privacy protection but it comes with a cost in terms of  utility \citep{wagh2021dp}. Then, adding the user-level DP  adversely affects the utility.

The root cause of these limitations is that the combination of sentence-level DP and user-level DP notions does not capture the correlation between sensitive entities and user membership in unifying notion of DP. Meanwhile, working with word-level embedded features under local DP introduces expensive model utility costs. 
Therefore, there is a demand for a unifying notion of DP and an optimal approach to protect both sensitive entities and user membership in training NLMs.

\subsection{UeDP Definition}

To preserve privacy for both users and sensitive entities in NLMs, we propose a new definition of user-entity adjacent databases, as follows: Two databases $D$ and $D'$ are user-entity adjacent if they differ in a single user and a single sensitive entity; that is, one user $u'$ and one sensitive entity $e'$ are present in one database (i.e., $D'$) and are absent in the other (i.e., $D$). 
Together with the absence of all sentences from the user $u'$ in $D$, all sentences (across users) consisting of the sensitive entity $e'$ are also absent in $D$. This is because one user can have multiple sentences, and one sensitive entity can exist in multiple sentences for training. The definition of our user-entity adjacent databases is presented as follows:
  \begin{mydef}{User-Entity Adjacent Databases.} Two databases $D$ and $D'$ are called user-entity adjacent if: $\|U-U'\|_1 \le 1$ and $\|E-E'\|_1  \le 1$, where $U$ and $E$ are the sets of users and sensitive entities in $D$, and $U'$ and $E'$ are the sets of users and sensitive entities in $D'$.
  \label{UeAB} 
 \end{mydef}

Given the user-entity adjacent databases, we present our UeDP in the following definition.

\begin{mydef}{$(\epsilon, \delta)$-UeDP.} A randomized algorithm $\mathcal{A}$ is  $(\epsilon, \delta)$-UeDP  if for all outcomes $ \mathcal{O}\subseteq Range(\mathcal{A})$ and for all user-entity adjacent databases $D$ and $D'$, we have: 
\begin{equation}
Pr[\mathcal{A}(D) = \mathcal{O}] \leq e^{\epsilon} Pr[\mathcal{A}(D') =  \mathcal{O}] + \delta 
\end{equation}
with a privacy budget $\epsilon$ and a broken probability $\delta$.
\label{UCDP} 
\end{mydef}



\section{Preserving UeDP in NLMs} 
\label{bound noise scale}

UeDP provides rigorous privacy protection to both users and sensitive entities; however, the practicability of UeDP preservation depends on the reliability of  sensitive entity detection from the training text data. In practice, misidentifying sensitive entities can introduce extra privacy risks.
In addition to addressing this challenge, we focus on bounding the sensitivity of an NLM under UeDP and addressing the trade-off between privacy loss and model utility.


\subsection{Misidentifying Sensitive Entities} 

Identifying all the sensitive entities typically requires intensive manual efforts \citep{yang2018automated}. We are aware of this issue in real-world applications. 
Fortunately, there are several ways to automatically identify sensitive entities in textual data, such as: 1) Using Named Entity Recognition (NER) \citep{sang2003introduction,derczynski2017results}; and 2) Using publicly available toolkits for detecting named entities or PII in text, e.g., spaCy \citep{honnibal2017spacy}, Stanza \citep{qi2020stanza}, and Microsoft Presidio\footnote{\url{https://microsoft.github.io/presidio/}}.
These approaches and toolkits are user-friendly and reliable to reduce manual efforts in identifying sensitive entities and information. We found that the results from spaCy cover over $94\%$ of sensitive information identified by Amazon Mechanical Turk (AMT) workers in a diverse set of datasets used in our experiments. 
More information about identifying sensitive entities is available in Appx.~\ref{Sensitive Entity Categories}. 

Although effective, the small error rate (i.e., $\approxeq 6\%$) from these techniques introduces a certain level of privacy risk, that means, some sensitive entities may be misidentified to be non-sensitive, and vice-versa. Classifying non-sensitive entities to be sensitive entities does not incur any extra privacy risk. Meanwhile, classifying one (or more than one) sensitive entity to be non-sensitive in a sentence introduces two issues, as follows: \textbf{(1)} There may be sensitive sentences  misidentified to become non-sensitive sentences. In order words, given a set of non-sensitive sentences detected by NER tools, we do not know which sentence is truly non-sensitive; and \textbf{(2)} Given a sensitive sentence $x$, some sensitive entities in $x$ may not be identified by NER tools. Preserving UeDP in NLMs by directly using the results of NER tools will expose these misidentifying sensitive sentences and entities unprotected. 


\subsection{Preserving UeDP}

To address the problem of sensitive entity misidentification in preserving UeDP, our key idea is: 

\textbf{(1)} Extending UeDP by considering each sentence, identified to be non-sensitive using NER tools \citep{honnibal2017spacy,qi2020stanza}, in the private training dataset as a single type of sensitive entity. \textit{We denote this extended set of sensitive entities as $S$}. The private dataset $D$ now consists of $U$ users and a (sufficient) set of sensitive entities $E \cup S$ that will be protected. 

\textbf{(2)} Upon forming the sufficient set of sensitive entities, we propose a two-step sampling approach to strictly preserve UeDP in NLMs. In our approach, at a training round $t$, we sample a set of users from $U$ and a set of sensitive entities from $E \cup S$. We use sentences in the training data of the sampled users consisting of the sampled sensitive entities to train NLMs. In this sampling approach: (i) If a sensitive sentence $x$ is not sampled for training, i.e., due to the fact that some sensitive entities in $x$ are not identified by NER tools, $x$ is not used for training at the round $t$; thus avoiding privacy risks exposed by $x$; and (ii) If the sensitive sentence $x$ is sampled for training, then the sensitive entities in $x$, which are not identified by NER tools, are protected since $x$ is protected under DP.

By covering all possible cases of sensitive entity misidentification, we strictly preserve UeDP without having additional privacy risks. The pseudo-code of our algorithm is in Alg.~\ref{alg1}.

At each iteration $t$, we  randomly sample $U^t$ users from $U$, $E^t$ detected sensitive entities from $E$, and $S^t$ extended sensitive entities from $S$, with sampling rates $q_u$, $q_e$, and $q_s$, respectively (Lines 8 and 10). Then, we use all sensitive sentences in $E^t_u \cup S^t_u$ consisting of the sensitive entities in $E^t$ and $S^t$ belonging to the selected users in $U^t$ for training. Like \citep{mcmahan2017learning}, we leverage the basic federated learning setting in \citep{mcmahan2016federated} to compute gradients of model parameters for a particular user, denoted as $\Delta^{t+1}_{u, \mathcal{E}}$ (Line 11). Here, we clip the per-user gradients so that its $l_2$-norm is bounded by a predefined gradient clipping bound $\beta$ (Lines 20 - 29). Next, a weighted-average estimator $f_{\mathcal{E}^+}$ is employed to compute the average gradient $\Delta^{t+1}$ using the clipped gradients $\Delta^{t+1}_{u, \mathcal{E}}$ gathered from all the selected users (Line 13). Finally, we add random Gaussian noise $\mathcal{N} (0, I \sigma^2)$ to the model update (Line 15).
During the training, the moments accountant $\mathcal{M}$  is used to compute the $T$ training steps’ privacy budget consumption (Lines 16 - 18).

To tighten the sensitivity bound, our weighted-average estimator $f_{\mathcal{E}^+}$ (Line 13) is as follows: 
\begin{align}
f_{\mathcal{E}^+}(U^{t}, E^{t}) = \frac{\sum_{u \in U^t}w_u \Delta_{u,\mathcal{E}}^{t+1} }{q_uW_u  (q_eW_e + q_{{s}}W_{{s}})} 
\label{estimator fc+}
\end{align} 
where $\Delta^{t+1}_{u,\mathcal{E}} = \sum_{e \in E^t_u} w_e \Delta_{u,e} + \sum_{s \in {S}^t_u} w_s \Delta_{u,s}$, and  $w_u$, $w_e$, and  $w_s \in [0,1]$ are weights associated with a user $u$, a detected sensitive entity $e$, and an extended sensitive entity $s$.


These weights capture the influence of a user and sensitive entities to the model outcome.  $\Delta_{u,e}$ and $\Delta_{u,s}$ are the parameter gradients computed using the sensitive entities $e \in E$ and $s \in S$. 
In addition, $W_u = \sum_{u\in U} w_u$, $W_e = \sum_{e \in E} w_e$, and $W_{{s}} = \sum_{s \in {S}} w_s$.

Since $\mathbb{E}[\sum_{e \in E^t_u} w_e + \sum_{s \in {S^t_u}} w_s] = q_eW_e + q_{{s}}W_{{s}}$, the estimator $f_{\mathcal{E}^+}$ is unbiased. 
The sensitivity of the estimator $\mathbb{S}(f_{\mathcal{E}^+})$ is computed as: $\mathbb{S}(f_{\mathcal{E}^+}) = \max_{u', e'} \lVert f_{\mathcal{E}^+}(\{U^t \cup u', (E^t \cup S^t) \cup e'\}) - f_{\mathcal{E}^+}(\{U^t, E^t \cup S^t\}) \rVert_2$.
$\mathbb{S}(f_{\mathcal{E}^+})$ is bounded in the following lemma.

\begin{lemma} If for all users $u$ we have $\lVert \Delta^{t+1}_{u,\mathcal{E}} \rVert_2 \leq \beta$, then $\mathbb{S}(f_{\mathcal{E}^+})  \leq \frac{(q_u|U| + 1) \max(w_u) \beta}{q_uW_u  (q_eW_e + q_{{s}}W_{{s}})}$.
\label{lemma2}
\end{lemma}

\begin{proof} If for all users  $\|\Delta^{t+1}_{u,\mathcal{E}} \rVert_2 \leq \beta$, then\\
\begin{align}
\small
& \nonumber  \mathbb{S}(f_{\mathcal{E}^+}) =
\frac{\sum_{u \in U^t \cup u'}w_u \Big(\sum_{e \in E^t_u} w_e (\sum_{\textit{s consists of }e} \Delta_{u,s})    \Big) }{q_uW_u  
(q_eW_e + q_{{s}}W_{{s}}) } \nonumber \\
& \quad \quad \quad  +  \frac{\sum_{u \in U^t \cup u'}w_u \Big( \sum_{s \in {S}^t_u} w_s \Delta_{u,s}\Big)    }{q_uW_u  
(q_eW_e + q_{{s}}W_{{s}}) } \nonumber \\
& \quad \quad \quad  + \frac{\sum_{u \in U^t \cup u'}w_u \Big[ w_{e'} (\sum_{\textit{s consists of }e'}\Delta_{u,s})   \Big] }{q_uW_u 
(q_eW_e + q_{{s}}W_{{s}}) } \nonumber\\
& \quad \quad \quad - \frac{\sum_{u \in U^t}w_u \Big(\sum_{e \in E^t_u} w_e (\sum_{\textit{s consists of }e} \Delta_{u,s}) \Big) }{q_uW_u  (q_eW_e + q_{{s}}W_{{s}})} \nonumber \\
& \quad \quad \quad - \frac{\sum_{u \in U^t}w_u \Big( \sum_{s \in {S}^t_u} w_s \Delta_{u,s}\Big) }{q_uW_u  (q_eW_e + q_{{s}}W_{{s}})} \nonumber \\
&\leq \frac{\sum_{u \in U^t \cup u'} [(w_u) \beta]}{q_uW_u  (q_eW_e + q_{{s}}W_{{s}})}  \leq \frac{(q_u|U| + 1) \max(w_u) \beta}{q_uW_u  (q_eW_e + q_{{s}}W_{{s}})}
\end{align}
Consequently, Lemma \ref{lemma2}  holds.
\end{proof}


Once the sensitivity of the estimator  $f_{\mathcal{E}^+}$ is bounded, we can add Gaussian noise scaled to the sensitivity $\mathbb{S}(f_{\mathcal{E}^+}) $  to obtain  a privacy guarantee. By applying Lemma \ref{lemma2}, the noise scale $\sigma$ becomes:
\begin{equation}
\sigma = z\mathbb{S}(f_{\mathcal{E}^+}) = \frac{ z (q_u|U| + 1) \max(w_u) \beta}{q_uW_u  (q_eW_e + q_{{s}}W_{{s}})}
\label{noise scale+}
\end{equation}

The noise scale $\sigma$ in Eq.~\ref{noise scale+} is tighter than the noise scale in existing works \citep{mcmahan2017learning,ramaswamy2020training} proportional to the number of sensitive entities used in the training process (i.e., $q_eW_e + q_{{s}}W_{{s}}$). Therefore, we can inject less noise into our model under the same privacy budget while improving our model utility.

\begin{algorithm2e}[t] 
\small
\caption{UeDP-Alg}
\begin{algorithmic}[1]\label{alg1}
\STATE \textbf{Input}: Dataset $D$, set of sensitive entities $E$, extended set of sensitive entities ${S}$, sampling rates $q_u$, $q_e$, and $q_{{s}}$, clipping bound $\beta$, a hyper-parameter $z$,  and number of iterations $T$
\STATE Initialize model $\theta^0$ and moments accountant $\mathcal{M}$  
\STATE $w_u \leftarrow \min (\frac{n_u}{\hat{w}_u}, 1)$ for all users $u$ 
\STATE $w_e \leftarrow \min (\frac{n_e}{\hat{w}_e}, 1)$ for all sensitive entities in $E$ 
\STATE $w_s \leftarrow \min (\frac{n_s}{\hat{w}_s}, 1)$ for all extended sensitive entities in ${S}$ \\
where  $n_u$, $n_e$, and
$n_s$ are 
the number of sentences in user $u$, the number of sentences containing sensitive entities $e \in E$, the number of sensitive entities in ${S}$, and $\hat{w}_u$, $\hat{w}_e$, and $\hat{w}_s$ are per-user sentence cap, per-entity sentence cap, and per-entity entity cap.
\STATE $W_u \leftarrow \sum_{u \in U} w_u$, $W_e \leftarrow \sum_{e \in E} w_e$,  $W_{{s}} \leftarrow \sum_{s \in {S}} w_s$ \\
\FOR {$t \in T$}
\STATE $U^t \leftarrow$ sample users with probability $q_u$
\FOR {each user $u \in U^t$}
\STATE $E_u^t \cup S_u^t \leftarrow$ sensitive entities (belonging to the user $u$) consisting of sensitive entities $E^t$ sampled from $E$ with probability $q_e$ and extended sensitive entities $S^t$ sampled from $S$ with probability $q_s$
\STATE $\Delta_{u,\mathcal{E}}^{t+1} \leftarrow \textbf{Local-Update}(u, E_u^t \cup S_u^t, \theta^t,  \text{ClipFn})$
\ENDFOR
\STATE $  \Delta^{t+1} \leftarrow \frac{\sum_{u \in U^t}w_u \Delta_{u,\mathcal{E}}^{t+1} }{q_uW_u  (q_eW_e + q_{{s}}W_{{s}})}$

\STATE $  \sigma \leftarrow \frac{ z (q_u|U| + 1) \max(w_u) \beta}{q_uW_u  (q_eW_e + q_{{s}}W_{{s}})}$ 

\STATE $\theta^{t+1} \leftarrow \theta^t + \Delta^{t+1} + \mathcal{N} (0, I \sigma^2)$
\STATE $\mathcal{M}$.\verb|accum_priv_spending(z)|
\ENDFOR 
\STATE print $\mathcal{M}$.\verb|get_priv_spent()|
\STATE \textbf{Output:} $(\epsilon,\delta)$-UeDP $\theta$, $\mathcal{M}$
\STATE \textbf{\text{Local-Update}}($u, E_u^t \cup S_u^t, \theta^0$, ClipFn):
\begin{ALC@g}
\STATE  $\theta \leftarrow \theta^0$
\STATE $\mathcal{B} \leftarrow $ $u$’s data split into size $B$ batches \\
\FOR{batch $b \in \mathcal{B}$}
     \STATE $\forall e \in E^t_u: \Delta_{u, e} \leftarrow \sum_{\textit{sentence s $_{(\in b)}$ consists of }e} \bigtriangledown l (\theta, s)$
      \STATE $\forall s \in S^t_u \cap b: \Delta_{u, s} \leftarrow \bigtriangledown l (\theta, s)$
     \STATE $\Delta_{u,\mathcal{E}} \leftarrow \sum_{e \in E_u^t} w_e \Delta_{u, e} + \sum_{s \in {S}^t_u} w_s \Delta_{u, s}$ \\
     \STATE $\theta \leftarrow \theta^0 - \eta \Delta_{u, \mathcal{E}}$
\ENDFOR
\STATE return  ClipFn($\theta - \theta^0$, $\beta$)
\end{ALC@g}
\STATE \textbf{\text{ClipFn}}($\Delta, \beta$): return $\pi(\Delta, \beta) \leftarrow \Delta \cdot \min \Big(1, \frac{\beta}{ \| \Delta \|} \Big) $
\end{algorithmic}
\end{algorithm2e}
\setlength{\textfloatsep}{5pt}

In extreme cases, that is also true: \textit{(1) $E$ is empty,}  which means there are no detected sensitive entities. Given a fixed set of training data, while $E$ is empty, $S$ becomes larger (i.e., covering the whole dataset), resulting in a larger value of $W_s$. Therefore, we obtain a larger value of $q_{{s}}W_{{s}}$ (with a pre-defined $q_s$), enabling us to reduce the noise scale under the same UeDP guarantee. That is an advantage compared with the naive approach that only uses detected sensitive entities $E$ in the training process (i.e., ignoring the term $q_sW_s$ in Eq. \ref{noise scale+}). If $E$ is empty, the naive approach will have no sentences for training; and \textit{(2) $S$ is empty,}  that is, every sentence in the data consists of at least one detected sensitive entity $e \in E$. 
Similarly, given a fixed set of training data, if $S$ is empty, then $E$ and $W_e$ become larger. It enables us to obtain a larger value of $q_{{e}}W_{{e}}$ (with a pre-defined $q_e$), which results in smaller noise scale while maintaining the high model utility.

\textit{UeDP Guarantee.} 
Given the bounded sensitivity of the estimator, the moments accountant $\mathcal{M}$ \citep{abadi2016deep} is used to get a tight bound on the total UeDP privacy consumption of $T$ steps of the Gaussian mechanism with the noise $\mathcal{N} (0, I \sigma^2)$ (Line 15).

\begin{theorem}
For the estimator $f_{\mathcal{E}^+}$, the moments accountant of the sampled Gaussian mechanism correctly computes  UeDP privacy loss with the scale  $z = \sigma/\mathbb{S}( f_{\mathcal{E}^+})$ for $T$ training steps.
\label{theorem1}
\end{theorem}

\begin{proof}
At each step, users, detected sensitive entities in $E$, and extended sensitive entities in $S$ are selected randomly with probabilities $q_u$, $q_e$, and $q_{s}$, respectively. For 
$f_{\mathcal{E}^+}$, if the $l_2$-norm of each user's gradient update, using the sampled sensitive entities in $E_u^t \cup S_u^t$, is bounded by $\mathbb{S}( f_{\mathcal{E}^+})$, then the moments accountant can be bounded by that of the sampled Gaussian mechanism with sensitivity $1$, the scale $z = \sigma / \mathbb{S}( f_{\mathcal{E}^+})$, and sampling probabilities $q_u$, $q_e$, and $q_{{s}}$. Thus, we can apply the composability as in Theorem 2.1 \citep{abadi2016deep} to correctly compute the UeDP privacy loss with the scale $z = \sigma / \mathbb{S}( f_{\mathcal{E}^+})$ for $T$ steps.
\end{proof} 



\section{Experimental Results}
\label{exp}

We conducted an extensive experiment, both in theory and on benchmark datasets, to shed light on understanding  1) the integrity of sensitive entity identification,  2) the interplay among the UeDP privacy budget $(\epsilon, \delta)$, different types of sensitive entities (i.e., organization, location, PII, and miscellaneous entities), and model utility, and 3) whether considering the extended set of sensitive entities $S$ will improve our model utility under the same UeDP protection. 

\begin{table*}[t] 
\caption{Breakdown of CONLL-2003, AG, and SEC datasets.} 
\label{tb1}
\begin{center}
\begin{tabular}{|c|c|c|c|c|c|c|c|c|c|c|c|c|}
 \hline
\textbf{Dataset}  & \textbf{$|\mathcal{V}|$}  & \textbf{$\#$ of sentences}    & \textbf{$\#$ of users}    & \multicolumn{5}{c|}{\textbf{$\#$ of sensitive sentences}}  \\
 \hline
 \multirow{2}*{CONLL-2003} & \multirow{2}*{8,882} & \multirow{2}*{14,040} & \multirow{2}*{946} & Org    & Loc    & Person & Misc  & All   \\
 \cline{5-9} &  & & &  5,187     &    5,433       & 4,406    & 3,438  & 11,176  \\
\hline
 \multirow{2}*{AG} & \multirow{2}*{30,000} & \multirow{2}*{112,000} & \multirow{2}*{7,536} & Org    & Loc    & GPE & PII   & All  \\
 \cline{5-9} &  & & &  58,177     &     39,988       & 18,506    & 42,683 & 67,157   \\
\hline
 \multirow{2}*{SEC} & \multirow{2}*{12,651} & \multirow{2}*{5,188} & \multirow{2}*{1,592} & Org    & Loc     & GPE & PII  & All    \\
 \cline{5-9} &  & & & 1,955      &    273       & 60    &  357 & 2,166  \\
\hline
\end{tabular} 
\end{center}
\end{table*} 

\begin{figure*}[ht] 
\centering
\subfloat[CONLL-2003 dataset]{\label{lm-gpta}\includegraphics[scale=0.33]{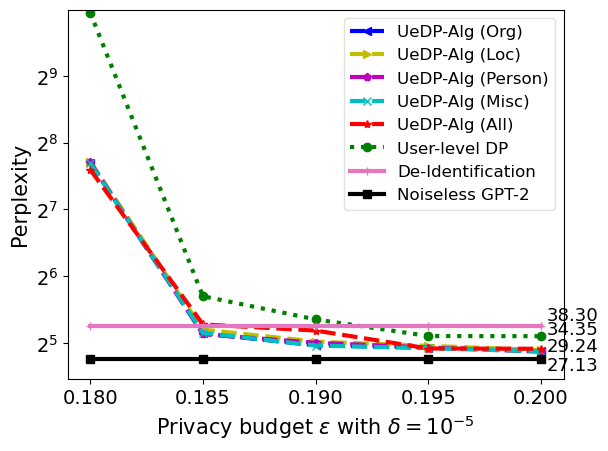}}\hfill
\subfloat[AG dataset]{\label{lm-gptb}\includegraphics[scale=0.33]{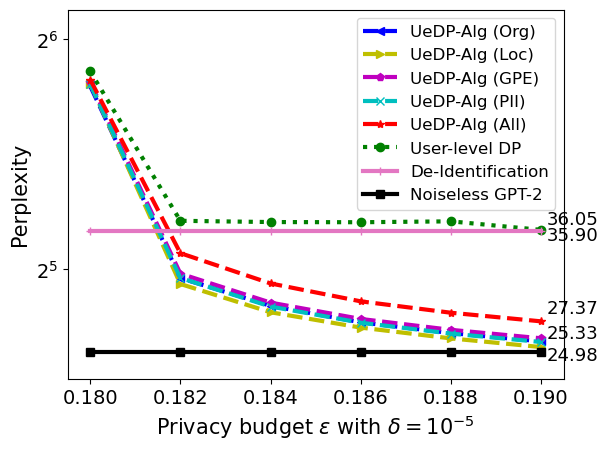}}
\hfill
\subfloat[SEC dataset]{\label{lm-gptc}\includegraphics[scale=0.33]{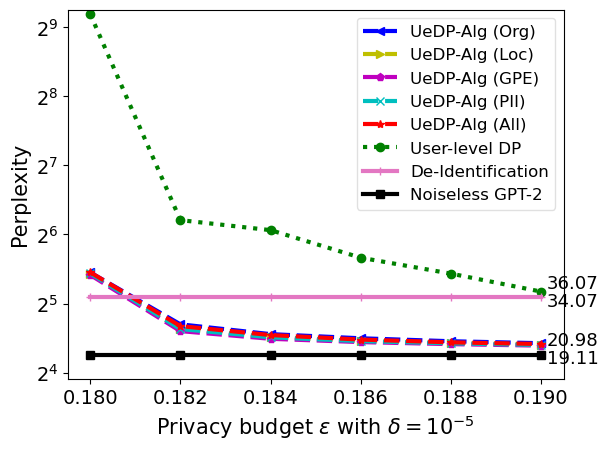}}\hfill
\caption{Next word prediction results using the GPT-2 model. (The lower the better)} \vspace{-7pt}
\label{lm-gpt}
\end{figure*}

\paragraph{Baseline Approaches}
We evaluate our \textbf{UeDP-Alg} in comparison with both noiseless and privacy-preserving mechanisms (either user level or entity level), including: \textbf{(1)} \textbf{User-level DP} \citep{mcmahan2017learning}, which is the state-of-the-art DP-preserving model closely related to our work; \textbf{(2)} \textbf{De-Identification}  \citep{dernoncourt2017identification}, which is considered as a strong baseline to protect privacy for sensitive entities. Although sensitive entities are masked to hide them in the training process, De-Identification does not offer formal privacy protection to either the data owners or sensitive entities; and \textbf{(3)} A \textbf{Noiseless} model, which is a language model trained without any privacy-preserving mechanisms.
In addition, we consider the naive approach, which is a variation of our algorithm, called \textbf{UeDP-Alg $f_{\mathcal{E}}$}. As a baseline, the estimator $f_{\mathcal{E}}$ is computed without taking the extended set of sensitive entities $S$ into account (Appx.~\ref{User-Entity Sensitivity Bound}, Supplementary$^4$). This is further used to comprehensively evaluate our proposed approach.  
In our experiment, our algorithms and baselines, i.e., UeDP-Alg, User-level DP, and De-Identification, are applied on the noiseless model in the training process.
As in the literature review \citep{asi2019element,lyu2020differentially}, there are no other appropriate DP-preserving baselines for UeDP protection.

\paragraph{Evaluation Tasks and Metrics} 
Our experiment considers two tasks: \textbf{(1)} next word prediction and \textbf{(2)} text classification. For the next word prediction, we employ the widely used perplexity \citep{mikolov2011extensions}. 
The smaller perplexity is, the better model is. For the text classification, we use the test error rate as in earlier work \citep{howard2018universal}. Test error rate implies prediction error on a test set, so it is $1$ - the test set's accuracy.
The lower the test error rate is, the better model is.



\paragraph{Data and Model Configuration} 
 For the reproducibility sake, all details about our datasets and data processing  are included in Appx.~\ref{hist of dataset} (Supplementary$^4$).
We carried out our experiment on three textual datasets, including the CONLL-2003 news dataset \citep{sang2003introduction}, AG's corpus of news articles\footnote{\url{http://groups.di.unipi.it/~gulli/AG_corpus_of_news_articles.html}},
 and our collected Security and Exchange Commision (SEC) financial contract dataset. 
 The data breakdown 
 is in Table~\ref{tb1}.
 
 For the next word prediction, we employ a GPT-2 model \citep{radford2019language}, which is one of the state-of-art text generation models. 
 To make the work easily reproducible, we use a version of the pretrained GPT-2 that has $12$-layer, $768$-hidden, $12$-heads, $117$M parameters, and then fine-tune with the aforementioned datasets as our Noiseless GPT-2 model.  For the text classification, we fine-tune a Noiseless BERT (i.e., BERT-Base-Uncased\footnote{\url{https://huggingface.co/transformers/pretrained_models.html}}) pre-trained model \citep{devlin2018bert}  that has $12$-layer, $768$-hidden, $12$-heads, and $110$M parameters with an additional softmax function on top of the BERT model. Adam optimizer is used with the learning rate is $10^{-5}$.  Gradient clipping bound $\beta=0.1$ and the scale $z =2.5$. The sampling rates for users, detected sensitive entities, and extended sensitive entities $q_u$, $q_e$, and $q_{{s}}$ are $0.05$, $0.5$, and $1.0$.

 To test the effectiveness and adaptability of our mechanism across models, we also conducted experiments with an AWD-LSTM model \citep{merity2017regularizing}, 
 which has a  much fewer parameters compared with GPT-2 and BERT.  In AWD-LSTM model, we use a three-layer LSTM model with $1,150$ units in the hidden layer and an embedding input layer of size $100$. Embedding weights are uniformly initialized in the interval $\lbrack - 0.1, 0.1 \rbrack$ with dimension $d=100$ and other weights are  initialized between $\lbrack - \frac{1}{\sqrt{H}}, \frac{1}{\sqrt{H}} \rbrack$, where $H$ is the size of all hidden layers.    
The values used for dropout on the embedding layer, the LSTM hidden-to-hidden matrix, and the final LSTM layer's output are $0.1$, $0.3$, and $0.5$, respectively. Gradient clipping bound $\beta=0.1$ and the scale $z =2$. The sampling rates $q_u$, $q_e$, and $q_{{s}}$ are $0.05$, $0.5$, and $1.0$ (note that $q_{{s}}$ is $0.6$ in the text classification task). SGD optimizer is used. In the text classification with the AG dataset, a softmax layer is applied on top of the AWD-LSTM with the output dimension is $4$, corresponding to four classes in the AG dataset. The same sets of sensitive entity categories are used for all models in the next word prediction and the text classification tasks.

\paragraph{Evaluation Results} To answer our evaluation questions, we conducted the following experiments:  \textbf{(1)} examining how the sensitive entities detected by the entity recognition spaCy \citep{honnibal2017spacy} covers the sensitive information clarified by AMT workers, \textbf{(2)} comparing estimators $f_{\mathcal{E}}$, $f_{\mathcal{E}^+}$, and User-level DP; \textbf{(3)} investigating the interplay between privacy budget and model utility; \textbf{(4)} studying the impacts of different sensitive entity categories in $E$ on the privacy budget and model utility; and \textbf{(5)} confirming our results in the text classification task. 

Our result is as follows:

\textbf{$\bullet$ Integrity of sensitive entities.} 
Our work utilizes spaCy \citep{honnibal2017spacy}, one of the state-of-the-art large-scale entity recognition systems, to identify sensitive entities for evaluation purposes on datasets that do not have ground-truth sensitive entities, including the AG and SEC datasets. For CONLL-2003, we consider the labels of four sensitive entity types (i.e., location, person, organization, and miscellaneous) from NER as the ground truth. To evaluate the integrity of identified sensitive entities, we conducted a clarification on AMT. We found that the results from spaCy cover over $94\%$ of sensitive information as identified by AMT workers. We recruited master-level AMT workers for a high quality of results, and we provided detailed guidance before AMT workers conducted the task. Each sentence was assigned to $3$ workers to mitigate bias and subjective views. Consequently, our experiments using the spaCy identified sensitive entities are solid.

\textbf{$\bullet$ Comparing Estimators $f_{\mathcal{E}}$, $f_{\mathcal{E}^+}$, and User-level DP.}
In this analysis, we set $q_u= 0.05$, $q_e = 0.5$, $q_{{s}} = 1$, $z= 2$, and compute privacy budget $\epsilon$ at $\delta = 10^{-5}$ (a typical value of $\delta$ in DP) as a function of the training steps $T$. 
Fig.~\ref{fig-uedp-e-eplus} shows curves of using different estimators and the User-level DP with all entities in CONLL-2003, AG, and SEC datasets. 

 Our UeDP-Alg with $f_{\mathcal{E}^+}$ achieves a notably tighter privacy budget compared with $f_{\mathcal{E}}$ and  the User-level DP in all scenarios in CONLL-2003, AG, and SEC datasets. The key reason is that 
 typically detected sensitive entities in $E$ appear rarely in a dataset compared with extended sensitive entities. Thus, using only sensitive entities in $E$ identified by the spaCy in training will cause information distortion, which can damage model utility and a loose privacy budget.

User-level DP consumes a much higher privacy budget $\epsilon$ compared with both of our estimators  $f_{\mathcal{E}^+}$ and $f_{\mathcal{E}}$. For instance, at $T=50$, the values of $\epsilon$ in all entities of $f_{\mathcal{E}^+}$ and $f_{\mathcal{E}}$, and the value of $\epsilon$ of the User-level DP in: \textbf{(1)} the CONLL-2003 dataset are $0.52$, $0.62$,  and $1.18$; \textbf{(2)} the AG dataset are  $0.50$, $0.75$,  and $1.48$; and \textbf{(3)} the SEC dataset are $0.40$, $0.71$, and $1.40$, respectively. 

Significantly, the privacy budget ($\epsilon$) gap between User-level DP, $f_{\mathcal{E}}$, and $f_{\mathcal{E}^+}$ is proportionally increased to the number of steps $T$. That means the more training steps $T$, the larger $\epsilon$
our model can save compared with  User-level DP. That is a promising result in the context that our model provides DP protection for both users and sensitive entities, compared with only protection for users in User-level DP. We observe a similar phenomenon on different sensitive categories. 

\begin{figure}[t] 
\centering
\hspace*{0.1in}
\subfloat[AG dataset]{\label{textclassa-bert}\includegraphics[scale=0.33]{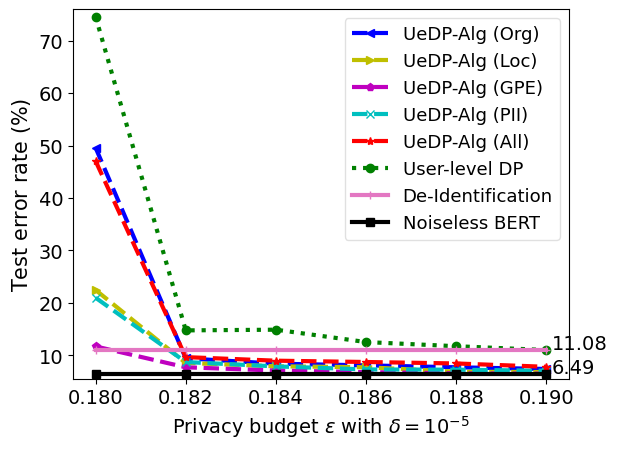}}\\
\subfloat[AG dataset with varying $q_{{s}}$]{\label{textclassb-bert}\includegraphics[scale=0.33]{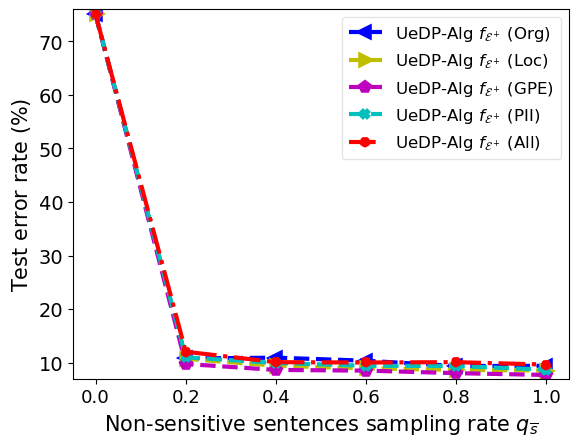}} 
\caption{Text classification results on the AG dataset using the BERT model. With $q_{{s}}=0.0$, the test error rate is $75\%$ in all cases. (The lower the better)} 
\label{textclass-bert}
\end{figure}
 
\textbf{$\bullet$ Privacy Budget $(\epsilon, \delta)$-UeDP and Model Utility.}   
From our theoretical analysis, $f_{\mathcal{E}^+}$ is better than the estimator $f_{\mathcal{E}}$. Therefore, for the sake of simplicity, we only consider UeDP-Alg $f_{\mathcal{E}^+}$ instead of showing results from both estimators. 
From now, UeDP-Alg is used to indicate the use of our estimator $f_{\mathcal{E}^+}$. 
Fig.~\ref{lm-gpt} illustrates the perplexity as a function of the privacy budget $\epsilon$ for an GPT-2 model trained on a variety of sensitive entity categories in UeDP, User-level DP, and De-Identification.
The noiseless GPT-2 (for the next word prediction) and  BERT (for the text classification) models are considered an upper-bound performance mechanism without offering any privacy protection.

In the CONLL-2003 dataset (Fig.~\ref{lm-gpta}), there are NER labels for person, location, organization, and miscellaneous entities; therefore, we choose these types as sensitive entity categories to protect in UeDP-Alg.  UeDP-Alg achieves a better perplexity compared with User-level DP under a tight privacy budget $\epsilon \in [0.18, 0.20]$. Also, from $\epsilon = 0.185$ (a tight privacy protection), our UeDP-Alg achieves a better perplexity than De-Identification. In fact,  at $\epsilon = 0.185$, our UeDP-Alg achieves $35.09$ for person, $35.34$ for organization,  $35.57$ for miscellaneous, and $36.79$ for location entities, compared with $52.01$ in User-level DP.  When spending more privacy budget ($\epsilon \geq 0.195$), both UeDP-Alg and User-level DP converge at a very competitive perplexity level, approaching the upper-bound Noiseless GPT-2. For instance, at $\epsilon = 0.20$, there are significant perplexity drops given UeDP-Alg and User-level DP mechanisms, i.e., our UeDP-Alg is $29.24$ for person, $29.35$ for miscellaneous, $29.58$ for organization, and $29.75$ for location  entities. Meanwhile, the perplexity values of User-level DP, De-Identification, and the noiseless GPT-2 model are $30.15$, $38.30$, and $27.13$.

\begin{figure*}
 \centering
\subfloat[CONLL-2003-all entities]{\label{d}\includegraphics[scale=0.33]{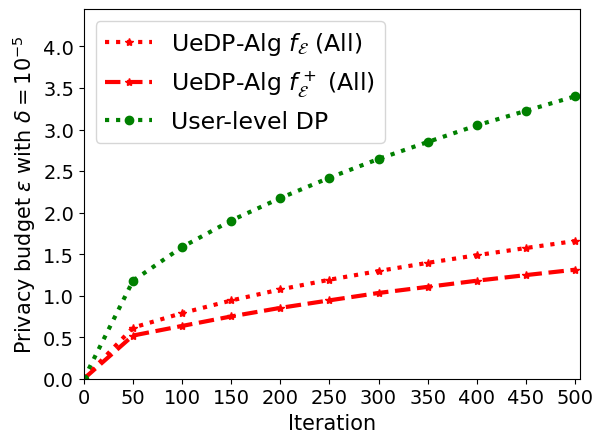}}\hfill
\subfloat[AG-all entities]{\label{e}\includegraphics[scale=0.33]{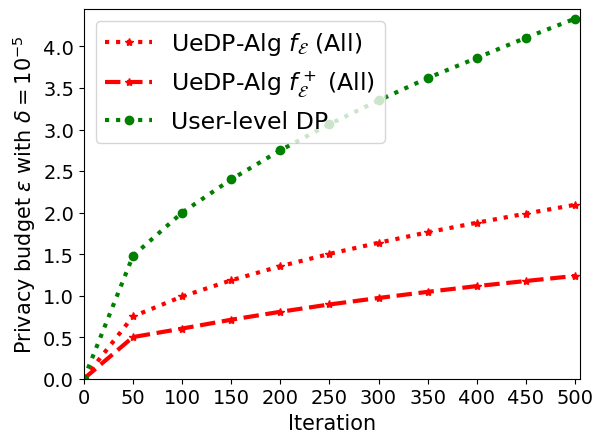}}\hfill
\subfloat[SEC-all entities]{\label{f}\includegraphics[scale=0.33]{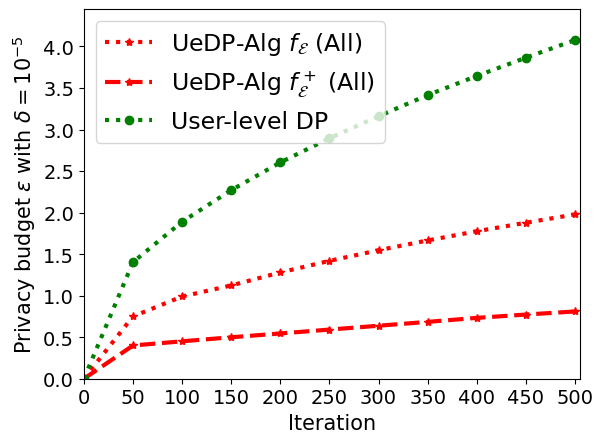}}\par 
\caption{Privacy budget of UeDP-Alg $f_{\mathcal{E}}$, UeDP-Alg $f_{\mathcal{E}^+}$, and User-level DP as a function of iterations in CONLL-2003, AG, and SEC datasets. UeDP-Alg $f_{\mathcal{E}^+}$ achieves a tighter privacy budget compared with UeDP-Alg $f_{\mathcal{E}}$ and User-level DP.  } \vspace{-2.5pt}
\label{fig-uedp-e-eplus}
\end{figure*}

 \begin{figure*} 
\centering
\subfloat[CONLL-2003 dataset]{\label{lma}\includegraphics[scale=0.33]{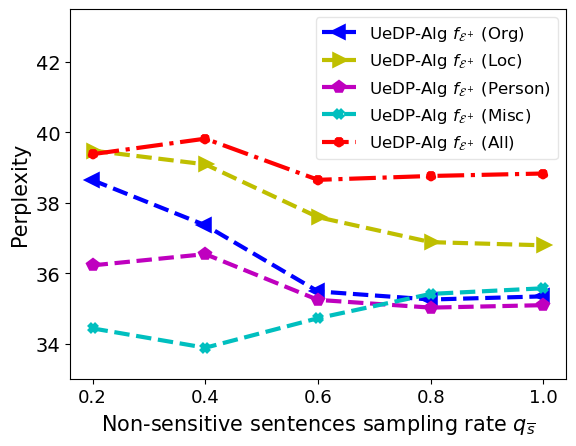}}\hfill
\subfloat[AG dataset]{\label{lmb}\includegraphics[scale=0.33]{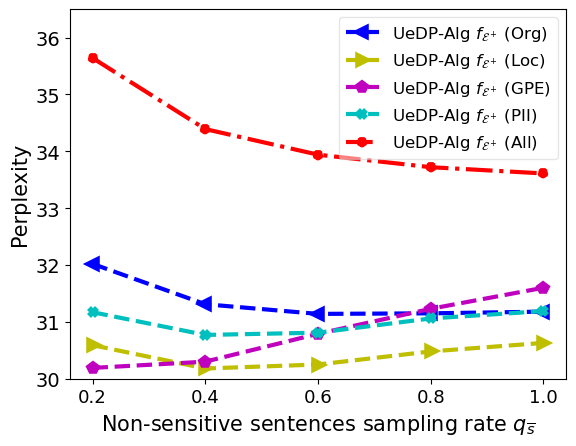}}\hfill
\subfloat[SEC dataset]{\label{lmc}\includegraphics[scale=0.33]{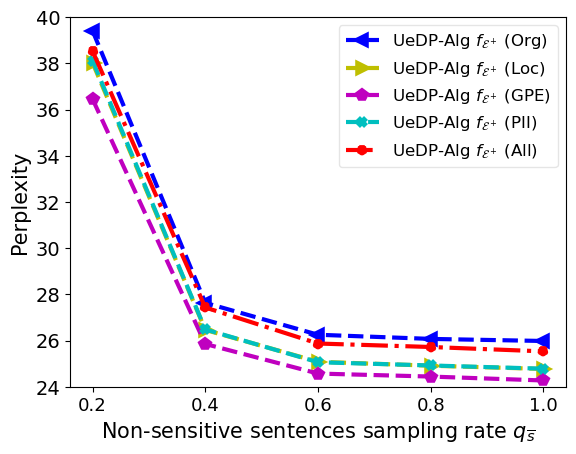}}\hfill
\caption{Next word prediction results using the GPT-2 model with varying extended sensitive entities sampling rate $q_{{s}}$ in training. (The lower the better)} 
\vspace{-2.5pt} 
\label{lm-gpt-qs}
\end{figure*}

 \begin{figure*} 
\centering
\subfloat[CONLL-2003 dataset]{\label{lma}\includegraphics[scale=0.33]{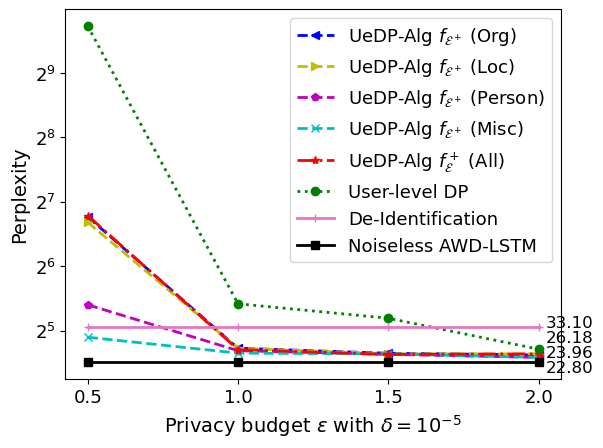}}\hfill
\subfloat[AG dataset]{\label{lmb}\includegraphics[scale=0.33]{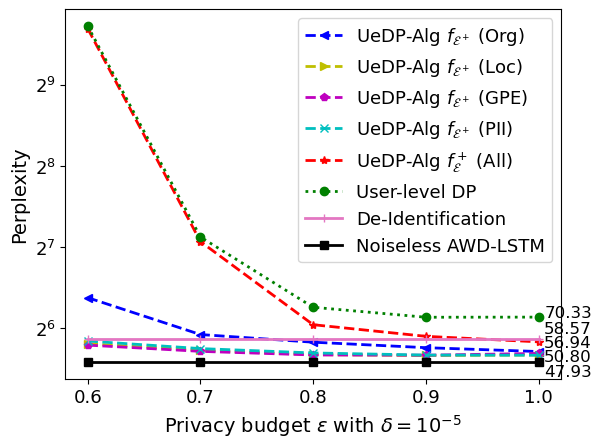}}\hfill
\subfloat[SEC dataset]{\label{lmc}\includegraphics[scale=0.33]{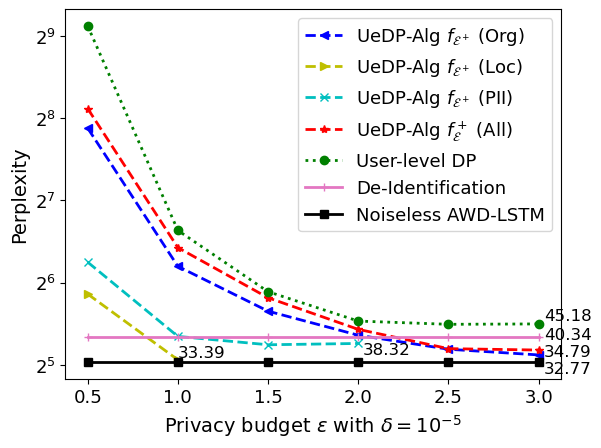}}\hfill
\caption{Next word prediction results using the AWD-LSTM model. (The lower the better)} \vspace{-2.5pt} 
\label{lm-awd-lstm}
\end{figure*}

  \begin{figure*}
 \centering
\subfloat[CONLL-2003]{\label{fig11d}\includegraphics[scale=0.33]{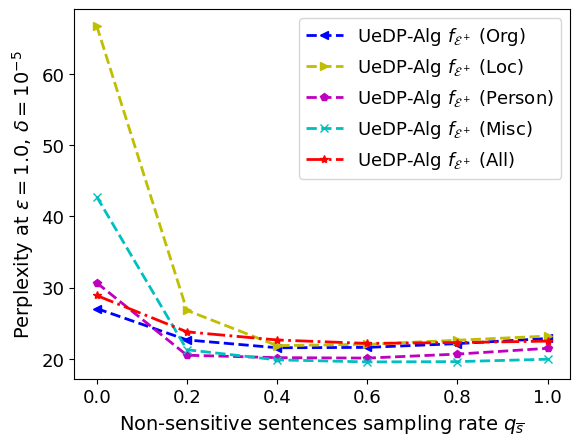}}\hfill
\subfloat[AG]{\label{fig11e}\includegraphics[scale=0.33]{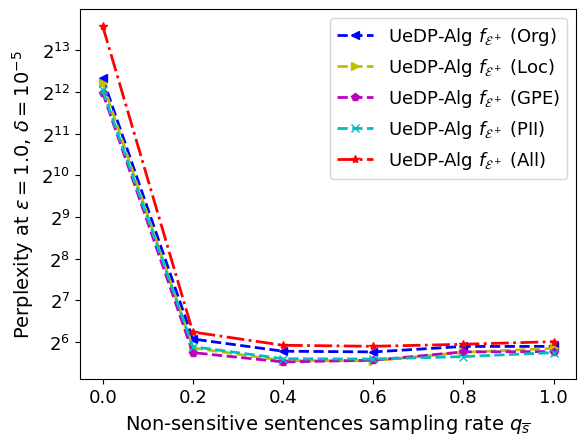}}\hfill
\subfloat[SEC]{\label{fig11f}\includegraphics[scale=0.33]{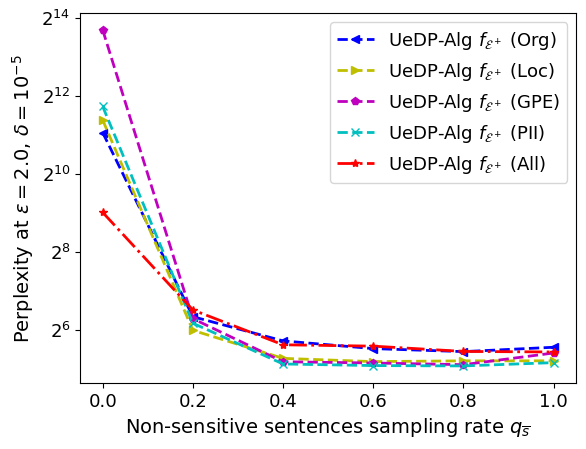}}\hfill
\caption{Next word prediction results using the AWD-LSTM model with varying extended sensitive entities sampling rate $q_{{s}}$ in training. (The lower the better)} 
\label{nonsens}
\end{figure*}

Results on AG and SEC datasets (Figs.~\ref{lm-gptb} and \ref{lm-gptc}) further strengthen our observations. In AG and SEC datasets, we applied spaCy to identify different sensitive entity categories, such as GPE, location, organization, and PII (i.e., person and location information). UeDP-Alg achieves better results compared with User-level DP in all considering sensitive entity categories and privacy budgets, and outperforms De-Identification in most cases. That is promising and consistent with our previous analysis. 
For instance, in the AG dataset, at $\epsilon = 0.19$, our UeDP-Alg achieves $25.33$ for location,  $25.72$ for PII, $25.77$ for organization, and $26.01$ for GPE  entities, compared with $36.05$ in User-level DP. De-Identification obtains $35.90$, and the upper bound result in the noiseless GPT-2 model is $24.98$. 
Similarly, in the SEC dataset (Fig.~\ref{lm-gptc}), at $\epsilon = 0.19$, UeDP-Alg achieves perplexity of $20.98$ in GPE, $21.12$ in PII, $21.22$ in location,  $21.50$ in organization, and $21.33$ in all entities, compared with $36.07$ in User-level DP, and $34.07$ in De-Identification. In AG and SEC datasets, at a tight privacy budget, i.e., $\epsilon =0.19$, our UeDP-Alg has better perplexity values than the De-Identification, approaching the noiseless GPT-2 model.

\textbf{$\bullet$ Sensitive Entity Categories.}
In all datasets (Figs.~\ref{lm-gpt} and \ref{fig-uedp-e-eplus2}, Appx.~\ref{diffsent}, Supplementary$^4$), \textit{the more sensitive sentences to protect, the higher the privacy budget is needed, and the lower performance the model achieves} (i.e., higher perplexity values). 
For instance, in the SEC dataset, the number of sensitive sentences in each category is as follows: $60$ in GPE, $273$ in location, $357$ in PII, $1,955$ in organization, and $2,166$ in all entities. 
After $500$ steps, the  values of $\epsilon$ are $0.19$ in GPE, $0.24$ in location, $0.26$ in PII, $0.73$ in organization, $0.81$ in all entities, and $4.08$ in User-level DP (Fig.~\ref{fig-uedp-e-eplus}). At $\epsilon = 0.18$ (Fig.~\ref{lm-gptc}), we obtain perplexity values of $42.63$ in GPE, $43.21$ in location, $43.30$ in PII, $43.70$ in organization, $43.77$ in all entities, and a $583.06$ in User-level DP.

\textbf{$\bullet$ Text classification.} Fig.~\ref{textclassa-bert} shows that our UeDP-Alg achieves lower test error rates in terms of text classification on the AG dataset than baseline approaches in most cases across different types of sensitive entities under a very tight UeDP protection ($\epsilon \in [0.18, 0.19]$). This is a promising result.
When $\epsilon$ is higher, the test error rates of both UeDP-Alg and User-level DP drop, approaching the noiseless BERT model's upper-bound result.

$\bullet$ \textbf{Extended Sensitive Entities.} To shed light into the impact the extended sensitive entity sampling rate $q_{{s}}$ on model utility under UeDP protection, we varied the value of $q_{{s}}$ from $0$ to $1$ in all datasets and tasks. Figs.~\ref{textclassb-bert},  \ref{lm-gpt-qs}, and \ref{nonsens} show that considering extended sensitive entities (i.e., $q_{{s}} > 0$) significantly improves model utility (i.e., perplexity or test error rate) compared with only considering sensitive entities $e \in E$ (i.e., $q_{{s}} = 0$). However, different tasks on different datasets may have different optimal values of $q_{{s}}$. 
This opens a new research question on how to theoretically approximate the optimal value of $q_{{s}}$.

Results on the AWD-LSTM model (Figs.~\ref{lm-awd-lstm} and \ref{nonsens}) further strengthen our observations. In our experiments, the AWD-LSTM model generally obtains comparable results with the GPT-2 model for next word prediction at a higher privacy budget range (i.e., $\epsilon \in [0.5, 3.0]$ in the AWD-LSTM model compared with $\epsilon \in [0.18, 0.2]$ in the GPT-2 model). This is because the GPT-2 model is pretrained on large-scale datasets, so that it is easily adapted to the idiosyncrasies of a target task (i.e., next word prediction) compared with the AWD-LSTM model trained from scratch.

\section{Conclusion and Future Work}
\label{con}

In this paper, we developed a novel notion of user-entity DP (UeDP), protecting users' participation information and sensitive entities in NLMs. By incorporating user and sensitive entity sampling in the training process, we addressed the trade-off between model utility and privacy loss with a tight bound of model sensitivity. Theoretical analysis and rigorous experiments show that  UeDP-Alg outperforms baselines in next word prediction and text classification under UeDP protection.
 
In practice, the list of sensitive entities and users can grow over time. Periodically updating the list of users and sensitive entities may incur extra privacy and computational cost. Therefore, we will focus on preserving UeDP 
given a growing list of users and sensitive entities in our future work.

\section*{Acknowledgement}
This work is partially supported by grants NSF IIS-2041096, NSF CNS-1935928, NSF CNS-1850094, and unrestricted gifts from Adobe System Inc.



\begin{footnotesize}
\fontsize{9pt}{9pt}\selectfont
\bibliographystyle{IEEEtran}
\bibliography{{IEEEfull}}
\end{footnotesize}

\newpage
\clearpage

\appendix

\subsection{Sensitive Entity Recognition and Tool-kits}
\label{Sensitive Entity Categories}

If a training set does not have sensitive entity indicators, we suggest several ways to identify sensitive entities in textual data, as follows.

\textbf{Using Named Entity Recognition (NER) datasets.} NER datasets \citep{sang2003introduction,derczynski2017results} refer to textual data in which entities in a text are labeled based on several predefined categories. NER typically makes it easy for individuals and systems to identify and understand the subject of the given text quickly. Therefore, extracted entities are critical and should be protected. 
For instance, in the CONLL-2003 dataset \citep{sang2003introduction}, there are four entity types, i.e., location, person, organization, and miscellaneous.  
    
\textbf{Using Publicly Available Tool-kits.} For textual datasets that do not have NER labels or sensitive entity indicators, there are publicly available tool-kits for detecting named entities or PII in text, for example, Spacy \citep{honnibal2017spacy}, Stanza \citep{qi2020stanza}, and Microsoft Presidio$^3$. Spacy and Stanza deploy pre-trained NER models based on statistical learning methods to identify eighteen categories of named entities, including person, nationality or religious groups, facility, etc. (Table \ref{tb3}). Microsoft Presidio is another toolbox for PII detectors and NER models based on Spacy and regular expression\footnote{\url{https://github.com/google/re2/}}. For instance, Spacy is used as a sensitive entity identification in Fig.~\ref{fig10} to detect ``David Johnson'' a person entity, ``Main'' a GPE entity, ``September 18'' a date entity, and ``Main Hospital'' an organization entity. 

We present descriptions of different sensitive entity categories in the CONLL-2003, AG, and SEC datasets in Table \ref{tb3}. The descriptions are from \citep{sang2003introduction} and spaCy, supporting eighteen different entity types. In the current work, we play with four different types and their combinations. Note that, in UeDP, providing the name of an algorithm and a sensitive entity means we consider that type of entity as sensitive entities in the training process. For instance, in Fig.~\ref{fig-uedp-e-eplus}, UeDP-Alg $f_{\mathcal{E}^+}$ (Org) means we use all organization entities as sensitive entities in the UeDP-Alg algorithm. ``All entities''  means all types of sensitive entities considered for the dataset are used. For example, ``all entities'' in the CONLL-2003 dataset means all person, location, organization, and miscellaneous entities are regarded as sensitive entities. Meanwhile, in the AG and SEC datasets, “all entities” means that all  organization, location, GPE, and PII entities are  considered sensitive entities. More entity types are also presented in Table \ref{tb3} so that users can have more choices when identifying sensitive entities.

\subsection{UeDP without Considering Extended Sensitive Entities}
\label{User-Entity Sensitivity Bound}

At each iteration $t$, we  randomly sample $U^t$ users from $U$ and $E^t$ sensitive entities from $E$, with sampling rates $q_u$ and $q_e$, respectively. Then, we use all sensitive sentences consisting of the sensitive entities in $E^t$ belonging to the selected users in $U^t$ for training. Like \citep{mcmahan2017learning}, we leverage the basic federated learning setting in \citep{mcmahan2016federated} to compute gradients of model parameters for a particular user, denoted as $\Delta^{t+1}_{u, \mathcal{E}}$. Here, we clip the per-user gradients so that its $l_2$-norm is bounded by a predefined gradient clipping bound $\beta$. Next, a weighted-average estimator $f_\mathcal{E}$ is employed to compute the average gradient $\Delta^{t+1}$ using the clipped gradients $\Delta^{t+1}_{u, \mathcal{E}}$ gathered from all the selected users. Finally, we add random Gaussian noise $\mathcal{N} (0, I \sigma^2)$ to the model update.
During the training, the moments accountant $\mathcal{M}$ is used to compute the $T$ training steps’ privacy budget consumption. 

In this process, we need to bound the sensitivity of the weighted-average estimator $f_\mathcal{E}$ for per-user gradients $\Delta^{t+1}_{u, \mathcal{E}}$. We first consider the following simple estimator, with both sampling rates $q_u$ for the user-level and $q_e$ for the sensitive entity-level: 
\begin{align}
\small
 & f_\mathcal{E}(U^t, E^t) = \frac{\sum_{u \in U^t}w_u \Delta_{u,\mathcal{E}}^{t+1} }{q_uW_u q_eW_e } \\
 & \textit{s.t. } \Delta^{t+1}_{u,\mathcal{E}} = \sum_{e \in E^t_u} w_e (\sum_{\textit{s consists of }e} \Delta_{u,s}) \nonumber 
\label{estimator fc}
\end{align} 
where $w_u$ and  $w_e \in [0,1]$ are weights associated with a user $u$ and with a sensitive entity $e$. These weights capture the influence of a user and a sensitive entity to the model outcome. 
$\Delta_{u,s}$ is the parameter gradients computed using a sensitive sentence $s$ consisting of the sensitive entity $e$. 
In addition, $W_u = \sum_{u} w_u$ and $W_e = \sum_{e} w_e$.

The estimator $f_{\mathcal{E}}$ is unbiased to the sampling process; since $\mathbb{E}[\sum_{u \in U^t} w_u] = q_uW_u$ and $\mathbb{E}[\sum_{e \in E^t_u} w_e] = q_eW_e$. The sensitivity of the estimator $f_\mathcal{E}$ can be computed as: $\mathbb{S}(f_{\mathcal{E}}) = \max_{u', e'} \lVert f_{\mathcal{E}}(\{U^t \cup u', E^t \cup e'\}) - f_{\mathcal{E}}(\{U^t, E^t\}) \rVert_2$,
where the added user $u'$ can have arbitrary data and $e'$ is an arbitrary sensitive entity.

Given that $\Delta^{t+1}_{u,\mathcal{E}}$ is $l_2(\beta)$-norm bounded, where $\beta$ is the radius of the norm ball by replacing $\Delta^{t+1}_{u,\mathcal{E}}$ with $\Delta^{t+1}_{u,\mathcal{E}} \cdot \min \Big(1, \frac{\beta}{ \| \Delta^{t+1}_{u,\mathcal{E}} \|} \Big)$, the sensitivity of  $\mathbb{S}(f_{\mathcal{E}})$ is also bounded.
\begin{lemma} If for all users $u$ we have $\lVert \Delta^{t+1}_{u,\mathcal{E}} \rVert_2 \leq \beta$, then $\mathbb{S}(f_{\mathcal{E}}) \leq \frac{(q_u|U| + 1) \max(w_u) \beta}{q_uW_u \times q_eW_e }$.
\label{lemma1} 
\end{lemma}

\begin{proof} If for all users $u$ we have $\lVert \Delta^{t+1}_{u,\mathcal{E}} \rVert_2 \leq \beta$, then $\mathbb{S}(f_{\mathcal{E}})$ 
\begin{align}
\nonumber
&= \frac{\sum_{u \in U^t \cup u'}w_u [(\sum_{e \in E^t} w_e (\sum_{s \in S^t_{ue}} \Delta_{u,s})) ] }{(q_uW_u \times q_eW_e)} \nonumber  \\
&+ \frac{\sum_{u \in U^t \cup u'}w_u [ w_{e'} (\sum_{s \in S^t_{ue'}}\Delta_{u,e'})]}{(q_uW_u \times q_eW_e)} \nonumber  \\
& - \frac{\sum_{u \in U^t }w_u [\sum_{e \in E^t} w_e (\sum_{s \in S^t_{ue}} \Delta_{u,s})]}{(q_uW_u \times q_eW_e)} \nonumber\\
&\leq \frac{\sum_{u \in U^t \cup u'} w_u \beta}{q_uW_u \times q_eW_e }   \leq \frac{(q_u|U| + 1) \max(w_u) \beta}{q_uW_u \times q_eW_e } 
\end{align}
Consequently, Lemma \ref{lemma1} holds.
\end{proof} 

By applying Lemma \ref{lemma1}, given a hyper-parameter $z$, the noise scale $\sigma$ for the estimator $f_\mathcal{E}$  is:
\begin{equation}
\sigma = z\mathbb{S}(f_{\mathcal{E}}) = \frac{ z (q_u|U| + 1) \max(w_u) \beta}{q_uW_u \times q_eW_e} 
\label{noise scale} 
\end{equation}


We show that this approach achieves $(\epsilon, \delta)$-UeDP, by applying the moments accountant $\mathcal{M}$  to bound the total privacy loss of $T$ steps of the Gaussian mechanism with the noise $\mathcal{N} (0, I \sigma^2)$ in Theorem \ref{theorem1}. However, this mechanism only uses sensitive entities detected by automatic toolkits to train the model ignoring a large number of extended sensitive entities. As a result, it introduces a loose sensitivity bound (Lemma \ref{lemma1}) and affects our model utility. 

\subsection{Datasets and Data Processing} 
\label{hist of dataset}

CONLL-2003  consists of Reuters news stories published between August 1996 and August 1997. CONLL-2003 is an NER dataset, where there are labels for four different types of named entities, including location, organization, person, and miscellaneous entities. These types of named entities are considered sensitive entities. In the CONLL-2003 dataset, there is no obvious user information; hence, we consider each document as a user consisting of multiple sentences in the next word prediction task. 

AG dataset is a collection of news articles gathered from more than $2,000$ news sources by ComeToMyHead academic news search engine\footnote{\url{http://newsengine.di.unipi.it/}}. It is categorized into four classes: world, sport, business, and science/technology. 
Similar to the CONLL-2003 dataset, there is no user information in AG. To imitate a user indicator, we randomly divide news into different users based on Gaussian distribution. There are no named entities; thus, we apply pre-trained Spacy to find named entities and PII in the dataset. We choose different types of these named entities to be sensitive entities: organization, GPE (i.e., countries, cities, and states), location, and PII entities. 

Our SEC dataset consists  contract clauses collected from contracts submitted in SEC filings\footnote{\url{https://www.sec.gov/edgar.shtml}}. Since the contracts can be associated with a company ID, we use the ID as a user indicator. Similar to the AG dataset, we consider organization, GPE, location, and PII entities as sensitive entities to protect.

In addition to the next word prediction, we conducted text classification on the AG dataset to further strengthen our observations. For text classification, the number of labels is not sufficient in the SEC dataset, and the labels do not exist in the CONLL-2003 dataset. Therefore, we do not utilize CONLL-2003 and SEC datasets for text classification in this study.

\begin{figure*}[t]
\centering
\subfloat[CONLL-2003 dataset]{\label{lma1}\includegraphics[scale=0.35]{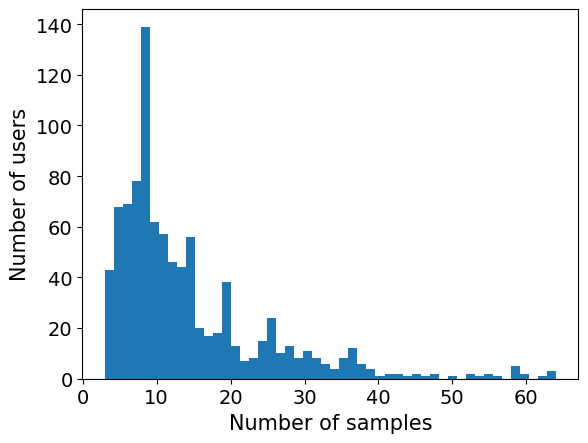}}\hfill
\subfloat[AG dataset]{\label{lmb1}\includegraphics[scale=0.35]{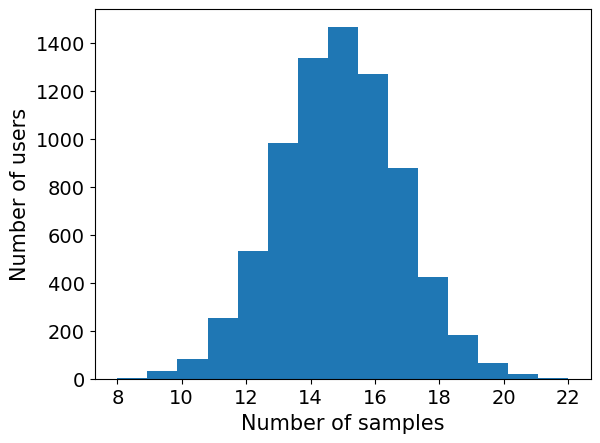}}\hfill
\subfloat[SEC dataset]{\label{lmc1}\includegraphics[scale=0.35]{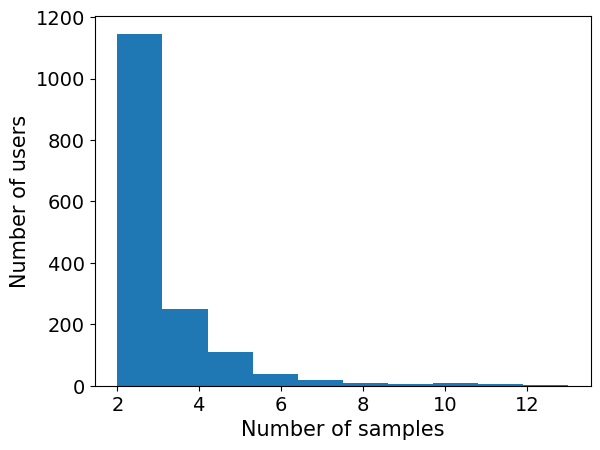}}  \par 
\caption{Distribution of users and sentences. }
\label{histogram}
\end{figure*}

For data preprocessing, we changed all words to lower-case and removed punctuation marks. 
Fig.~\ref{histogram} shows the distribution of the number of users and sentences in the CONLL-2003, AG, and SEC datasets. In the CONLL-2003 dataset, there is no obvious user information; hence, we consider each document as a user consisting of multiple sentences. Like the CONLL-2003 dataset, in the AG dataset, there is no user information. Therefore, to imitate a user indicator, we randomly divide news into different users. The number of sentences per user follows a Gaussian distribution $\mathcal{N}(15, 2^2)$, i.e., there are $15$ sentences per user on average, and the standard deviation is $2$ sentences.  
In the SEC dataset, since the contracts can be associated with a company ID, we use the ID as a user indicator. The document related to the ID is considered to be that user's data. 

\subsection{Revisiting Word-level LDP Analysis in \citep{lyu2020differentially}}
\label{lyu2020differentially}
This section aims at revisiting privacy protection in \citep{lyu2020differentially} and describes a  privacy accumulation issue over the embedding dimension. Then, we revise Theorems 1 and 2 in \citep{lyu2020differentially} and compare them with our approaches.

In \citep{lyu2020differentially}, the authors aim at preserving the privacy of the extracted test representation from users  while maintaining the good performance of the classifier, which is trained at a server by the data collected from users. To achieve the goal, they consider a word-level DP, that is, two inputs  $x$ and $x'$ are adjacent if they differ by at most 1 word. Additionally, they introduce a DP noise layer $r$ after a predefined feature extractor $f(x)$. To train a robust classifier at the server, they add the same level of noise as the test phase in the training process and optimize the classifier by minimizing the loss function as follows:
\begin{equation}
    \mathcal{L}(x,y) = \mathcal{X} (C(f(x) + r), y)
\end{equation}
where $C$ is the classifier, $y$ is the true label, and $\mathcal{X}$ is the cross entropy loss function. 

The Laplace noise layer $r$ is injected into the embedding $f(x)$ in which its coordinates  $r = \{r_1, r_2, \ldots, r_k \}$ are random variables drawn from the Laplace distribution defined by $Lap(b)$ with $b=\frac{\Delta_f}{\epsilon}$, $\epsilon$ is the privacy budget, and $\Delta_f$ is the sensitivity of the extracted representation. Here, $k$ is the dimension of $f(x)$.

Algorithm \ref{lyu} describes how to derive DP-preserving representation from the feature extractor $f$. Note that $x_s$ in the Algorithm \ref{lyu} is a sentence (equivalent to $x$ in our notation), which is considered to be sensitive and needs to be protected.

\textbf{Revisting Theorems 1 and 2 in \citep{lyu2020differentially}.} In the paper, the  authors consider adjacent sentences differing by one word. Changing  one word in $x$ may change the entire embedding vector $f(x)$. Each element of $f(x)$ is normalized into the range $[0,1]$ (Line 5, Algorithm \ref{lyu}), hence each element sensitivity of  $f(x)$ is $\Delta_f = 1$, the noise is $Lap(\Delta_f / \epsilon)$. Therefore, each element of the embedding  $f(x)$ consumes a privacy budget $\epsilon$. Since the $k$ elements of the embedding are derived from a single sensitive input $x$, applying the LDP mechanism $\mathcal{A}(.)$, i.e., $Lap(b)$, $k$ times will consume the privacy budget $k \times \epsilon$. This follows the composition property in DP. 
Note that the $k$ elements cannot be treated by using the parallel property in DP \citep{CompositionProp}, since all of them are derived from a single (data) input $x$, NOT from $k$ different inputs ($k$ different data samples). Consequently, the privacy guarantees in Theorems $1$ and $2$ of \citep{lyu2020differentially} is $k\epsilon$-DP, instead of $\epsilon$-DP as reported.  



In their experimental results, e.g., Table $2$ of \citep{lyu2020differentially}, the approach could achieve almost the same (and even better) model utility with noiseless model given the extremely low $\epsilon = 0.05$ using BERT embeddings. As our analysis, the privacy budget in Theorems $1$ and $2$ is $k \epsilon$, instead of $\epsilon$. Therefore, the proper privacy budget is at least $0.05 \times 768 = 38.4$. Similar results were reported through out the all in experiments. With this high value of the privacy budget, the word-level DP in \citep{lyu2020differentially} provides loose privacy protection.


 \balance
\textbf{Revisting Element-level DP in \citep{lyu2020differentially}.} During our discussion with the authors of \citep{lyu2020differentially}, the authors mentioned that their approach preserves a new notion of $(\epsilon, 0)$-element-level DP, i.e., two embeddings differ from one element, instead of a word-level DP. However, for the element-DP to hold, all the elements in the embedding $f(x)$ must be independent from each other, that is, changing one element will not result in changing any other element. If changing one element results in changing all the remaining elements, then element-DP will be suffered from the dimension of the embedding by following group privacy. In the current approach, changing one element means there is a change in the input data $x$ to occur. Equivalently, using BERT, any change in the input data $x$ will result in changing the whole embedding (all elements). Therefore, the condition of two neighboring embeddings only differing in only one element does NOT hold in theory and practice. Consequently, the introduced element-level DP does NOT hold at the level of $(\epsilon, 0)$-DP.

\begin{algorithm}[t] 
\caption{Differentially Private Neural Representation (DPNR) \citep{lyu2020differentially}}\label{lyu}
\begin{algorithmic}[1]
\STATE  \textbf{Input}: Each sensitive input $x_s \in \mathbb{R}^d$, feature extractor $f$
\STATE \textbf{Parameters}: Dropout vector $I_n \in \{0,1 \}^d$
\STATE Word dropout: $\tilde{x}_s \leftarrow x_s \odot I_n$, where $\odot$ performs a word-wise multiplication.
\STATE Extraction: $x_r \leftarrow f(\tilde{x}_s)$
\STATE Normalization: $x_r \leftarrow x_r - \min(x_r)/ ( \max(x_r) - \min(x_r) )$
\STATE Perturbation: $\hat{x_r} \leftarrow x_r + r$, $r_i \sim Lap(b)$
\STATE  \textbf{Output}: Perturbed representation $\hat{x_r}$.
\end{algorithmic} 
\end{algorithm} \setlength{\textfloatsep}{10pt}

\textbf{Our revising Theorems 1 and 2 in \citep{lyu2020differentially}.} Based upon our analysis, we introduce revised versions of the Theorems $1$ and $2$ in \citep{lyu2020differentially}, as follows.

\begin{theorem}{\textbf{Revised Theorem 1 in \citep{lyu2020differentially}.}}
Let the entries of the noise vector $r$ be drawn from $Lap(b)$ with $b = \frac{\Delta_f}{\epsilon}$. The Algorithm \ref{lyu} is $k \epsilon$-word-level DP, where $k$ is dimension of the embedding $f(x)$. 
\label{theorem1lyu}
\end{theorem}
\begin{proof}
Each element of the embedding $f$ is bounded in $[0,1]$, so $\Delta_f =1 $ for each element. By adding random noise variables drawn from the Laplace $Lap(b)$ with $b = \frac{\Delta_f}{\epsilon}$ into each element of $f$, each element consumes $\epsilon /k$ privacy budget. Since the $k$ elements of the embedding are derived from a single sensitive input $x$, applying the mechanism $Lap(b)$ $k$ times on the $k$ elements will consume the privacy budget $k \epsilon$. Therefore, the Algorithm \ref{lyu} is $k \epsilon$-word-level DP. 
\end{proof}

\begin{theorem}
Given an input $x \in D$, suppose $\mathcal{A}(x) = f(x) + r$ is $k \epsilon$-word-level DP, let $I_n$ with dropout rate $\mu$ be applied to $x$: $\tilde{x} = x \odot I_n$, then $\mathcal{A}(\tilde{x})$ is $\epsilon'$-word level-DP, where $\epsilon' = \ln [(1-\mu) \exp(k\epsilon) + \mu]$.
\label{theorem2lyu}
\end{theorem}

\begin{proof}
Suppose there are two adjacent inputs $x_1$ and $x_2$ that differ only in the $i$-th coordinate (word), say $x_{1i}=v$, $x_{2i} \neq v$. For arbitrary binary vector $I_n$, after dropout, $\tilde{x}_1 = x_1 \odot I_n$,  $\tilde{x}_2 = x_2 \odot I_n$, there are two possible cases, i.e., $I_{ni} =0$ and $I_{ni} =1$. 

If $I_{ni} =0$: Since $x_1$ and $x_2$ differ only in $i$-th coordinate, after dropout $\tilde{x}_{1i} = \tilde{x}_{2i}=0$, hence $x_1 \odot I_n = x_2 \odot I_n$. Then $Pr\{\mathcal{A}(x_1 \odot I_n) = S \} =Pr\{\mathcal{A}(x_2 \odot I_n) = S \} $.

If  $I_{ni} =1$: Since $x_1$ and $x_2$ differ only in $i$-th coordinate, after dropout $\tilde{x}_{1i} = v$, and  $\tilde{x}_{2i} \neq v$. Since $\mathcal{A} (x)$ is $k\epsilon$-word level-DP, then $Pr\{\mathcal{A}(x_1 \odot I_n) = S \} \le \exp(k\epsilon)Pr\{\mathcal{A}(x_2 \odot I_n) = S \} $. 

Combining these two cases, and $Pr [I_{ni} = 0] = \mu$, we have:
\begin{align}
\nonumber  & Pr\{\mathcal{A}(x_1 \odot I_n) = S \} \\
\nonumber &= \mu Pr\{\mathcal{A}(x_1 \odot I_n) = S \} + (1-\mu) Pr\{\mathcal{A}(x_1 \odot I_n) = S \}\\
\nonumber  &\le  \mu Pr\{\mathcal{A}(x_2 \odot I_n) = S \} \\
\nonumber  & + (1-\mu) \exp(k\epsilon) Pr\{\mathcal{A}(x_2 \odot I_n) = S \}\\
\nonumber  &= [(1-\mu) \exp(k\epsilon) + \mu ]Pr\{\mathcal{A}(x_2 \odot I_n) = S \}\\
   &= \exp \Big( \ln [(1-\mu) \exp(k\epsilon) + \mu ] \Big) Pr\{\mathcal{A}(x_2 \odot I_n) = S \}
\end{align}

Therefore, after dropout, the privacy budget is $\epsilon' = \ln [(1-\mu) \exp(k\epsilon) + \mu]$. 
\end{proof}

\textbf{Comparison with UeDP.} Apart from the privacy accumulation over the embedding dimension, in \citep{lyu2020differentially}, during training the model, the Laplace or Gaussian noise is drawn at every training iteration. Therefore, the model accesses the raw data at every iteration. As a result, the privacy budget at the training phase is accumulated over the number of training iterations, which can be a large number causing an exploded privacy budget in training. \citep{lyu2020differentially} focuses on protecting privacy at the inference time and use the noise in the training phase to obtain a more robust model without considering training data privacy. This is different from our goal to protect users and sensitive entities of training data, which is a more challenging task. Our UeDP-preserving model can be deployed to the end-users for a direct use in the inference phase, without demanding that the end-users send their data embedding to our server; therefore offering a more rigorous privacy protection and better usability. In addition to this, our approach offers more rigorous DP budget bounds compared with the DPNR algorithm in \citep{lyu2020differentially}, since DPNR consumes large DP budgets that is proportional to the commonly large dimension of the embedding $k$.

\subsection{Supplemental Experimental Results}
\label{diffsent}

\vspace{-10pt}

\begin{figure*} 
 \centering
\subfloat[CONLL-2003-organization entities]{\label{a}\includegraphics[scale=0.3]{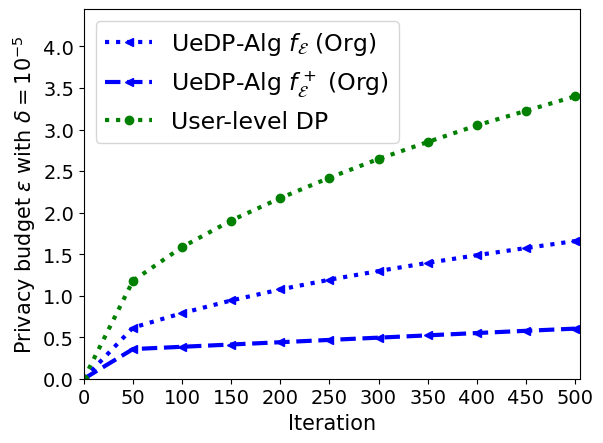}}\hfill
\subfloat[AG-organization entities]{\label{b}\includegraphics[scale=0.3]{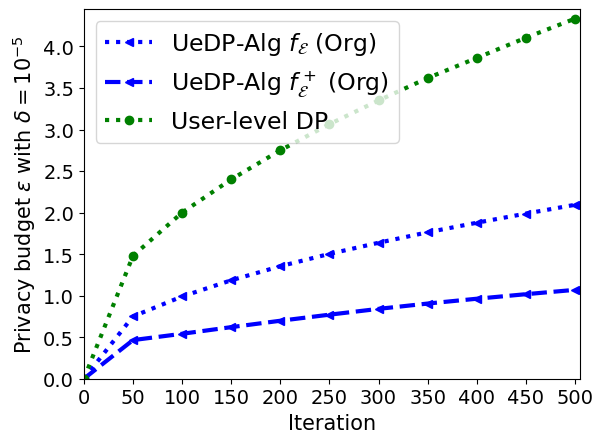}}\hfill
\subfloat[SEC-organization entities]{\label{c}\includegraphics[scale=0.3]{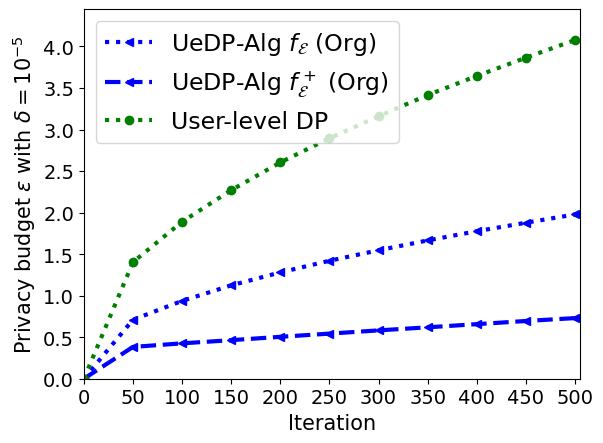}} \vspace{-10pt} \hfill
\subfloat[CONLL-2003-location entities]{\label{a}\includegraphics[scale=0.3]{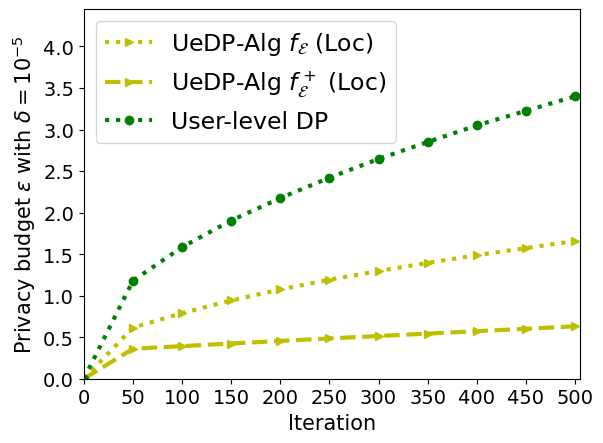}}\hfill
\subfloat[AG-location entities]{\label{b}\includegraphics[scale=0.3]{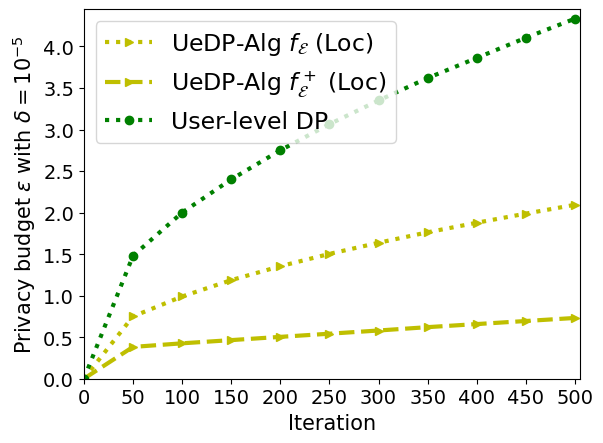}}\hfill
\subfloat[SEC-location entities]{\label{c}\includegraphics[scale=0.3]{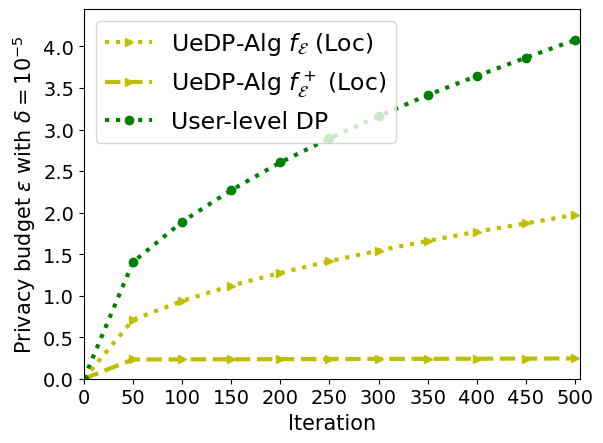}} \vspace{-10pt} \hfill
\subfloat[CONLL-2003-person entities]{\label{a}\includegraphics[scale=0.3]{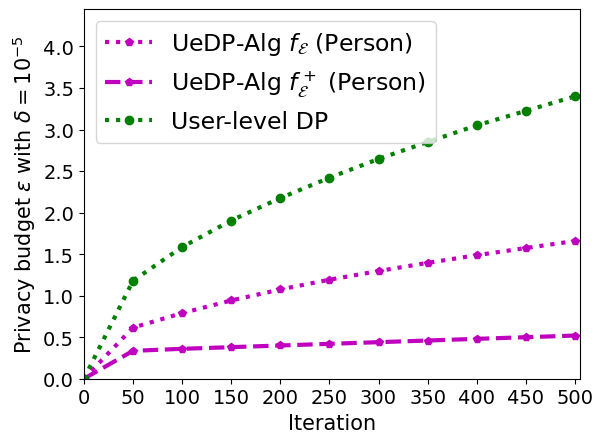}}\hfill
\subfloat[AG-GPE entities]{\label{b}\includegraphics[scale=0.3]{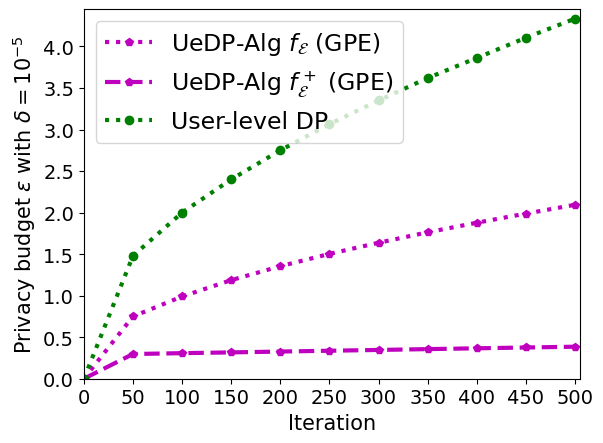}}\hfill
\subfloat[SEC-GPE entities]{\label{c}\includegraphics[scale=0.3]{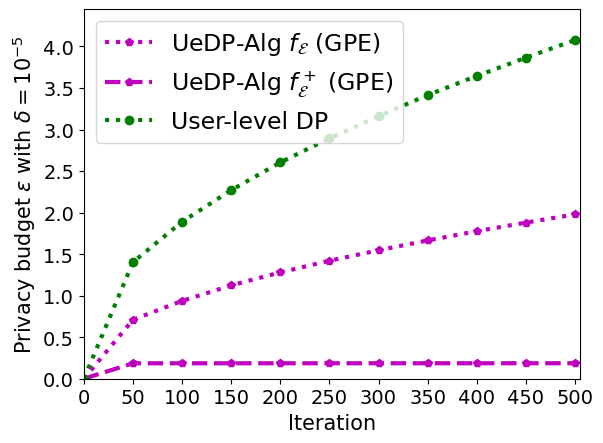}} \vspace{-10pt} \hfill
\subfloat[CONLL-2003-miscellanous entities]{\label{a}\includegraphics[scale=0.3]{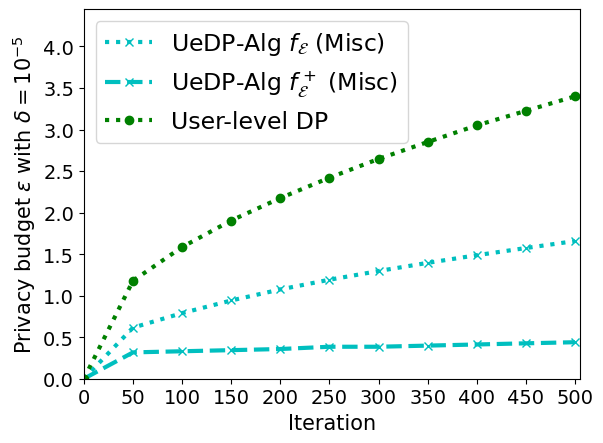}}\hfill
\subfloat[AG-PII entities]{\label{b}\includegraphics[scale=0.3]{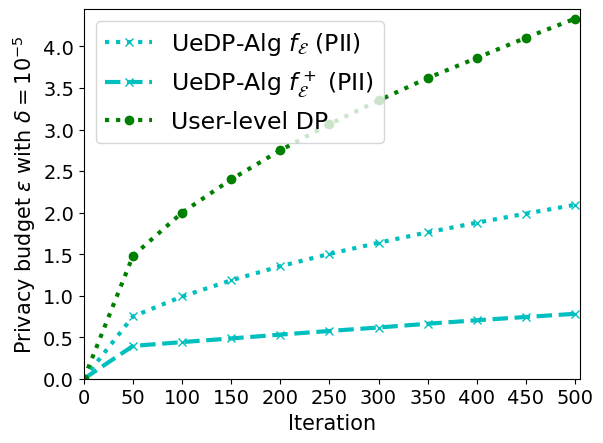}}\hfill
\subfloat[SEC-PII entities]{\label{c}\includegraphics[scale=0.3]{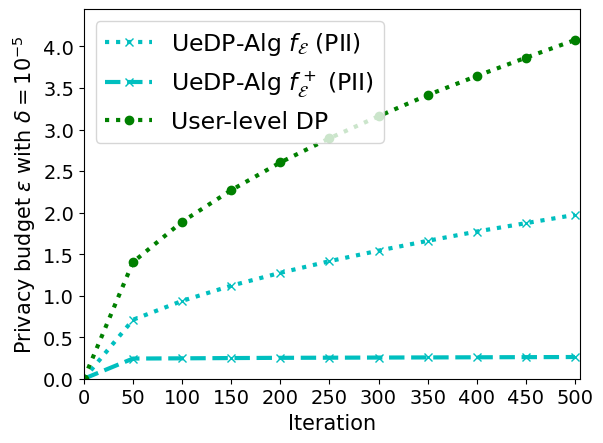}}  \hfill 
\caption{Privacy budget of UeDP-Alg $f_{\mathcal{E}}$, UeDP-Alg $f_{\mathcal{E}^+}$, and User-level DP as a function of iterations in CONLL-2003, AG, and SEC datasets. UeDP-Alg $f_{\mathcal{E}^+}$ achieves a tighter privacy budget compared with UeDP-Alg $f_{\mathcal{E}}$ and User-level DP.  }
\label{fig-uedp-e-eplus2}
\end{figure*}

\end{document}